\documentclass[nonacm,sigplan]{acmart}
\AtBeginDocument{%
  }

\setcopyright{none} 
\copyrightyear{2025}
\acmYear{2025}
\acmDOI{XXXXXXX.XXXXXXX}

\acmConference[MICRO 2025]{The 58th IEEE/ACM International Symposium on Microarchitecture}{October 18--22, 2025}{Seoul, Korea}
\acmISBN{978-X-XXXX-XXXX-X/XX/XX}

\newcommand{\frameworkname}{FluxTrap}
\usepackage{float}

\usepackage{stfloats}
\usepackage{amsmath,mathtools, amssymb, amsthm}

\usepackage[normalem]{ulem}

\usepackage{xcolor}
\usepackage{color}
\definecolor{codegreen}{rgb}{0,0.6,0}
\definecolor{codegray}{rgb}{0.5,0.5,0.5}
\definecolor{codepurple}{rgb}{0.58,0,0.82}
\definecolor{backcolour}{rgb}{0.95,0.95,0.92}
\definecolor{textblue}{rgb}{.2,.2,.7}
\definecolor{textred}{rgb}{0.54,0,0}
\definecolor{textgreen}{rgb}{0,0.43,0}
\definecolor{codered}{rgb}{201,72,12}
\usepackage[T1]{fontenc}
\usepackage[scaled=0.85]{beramono} 
\usepackage{listings}
\usepackage{multirow}
\usepackage{multicol}
\usepackage{url}
\usepackage[ruled,vlined,linesnumbered]{algorithm2e}
\usepackage{multirow}
\usepackage{tikz}
\usepackage{xcolor}
\usepackage{ctable} % for \specialrule command

\begin{document}

%\title{FluxTrap: Optimized Compilation for Efficient Qubit Routing on 2D Grid Ion-Trap Machines}
% \title{FluxTrap: Compiler Optimization for 2D Trapped-Ion Quantum Machines with SIMD Abstraction}
\title{TrapSIMD: SIMD-Aware Compiler Optimization for 2D Trapped-Ion Quantum Machines}
% \title{FluxTrap: Compiler Optimization for 2D Trapped-Ion Quantum Machines with SIMD Abstraction}
%\title{}
% \subtitle{\normalsize{MICRO 2025 Submission
%     \textbf{\#86} -- Confidential Draft -- Do NOT Distribute!!}}
%\author{\normalsize{ISCA 2025 Submission
 %   \textbf{\#NaN} -- Confidential Draft -- Do NOT Distribute!!}}



%%%%%% -- PAPER CONTENT STARTS-- %%%%%%%%
\author{Jixuan Ruan\textsuperscript{*}}
\email{j3ruan@ucsd.edu}
\affiliation{
    \institution{University of California}
    \city{San Diego}
    \country{USA}
}

\author{Hezi Zhang\textsuperscript{*}}
\email{hez019@ucsd.edu}
\affiliation{
    \institution{University of California}
    \city{San Diego}
    \country{USA}
}

\author{Xiang Fang}
\email{x8fang@ucsd.edu}
\affiliation{
    \institution{University of California}
    \city{San Diego}
    \country{USA}
}

\author{Ang Li}
\email{ang.li@pnnl.gov}
\affiliation{
    \institution{Pacific Northwest National Laboratory}
    \city{Richland}
    \country{USA}
}

\author{Wesley C. Campbell}
\email{wes@physics.ucla.edu}
\affiliation{
    \institution{University of California}
    \city{Los Angeles}
    \country{USA}
}

\author{Eric Hudson}
\email{eric.hudson@ucla.edu}
\affiliation{
    \institution{University of California}
    \city{Los Angeles}
    \country{USA}
}

\author{David Hayes}
\email{david.hayes@quantinuum.com}
\affiliation{
    \institution{Quantinuum}
    \city{Broomfield}
    \country{USA}
}

\author{Hartmut Haeffner}
\email{hhaeffner@berkeley.edu}
\affiliation{
    \institution{University of California}
    \city{Berkeley}
    \country{USA}
}

\author{Travis Humble}
\email{humblets@ornl.gov}
\affiliation{
    \institution{Oak Ridge National Laboratory}
    \city{Oak Ridge}
    \country{USA}
}

\author{Jens Palsberg}
\email{palsberg@cs.ucla.edu}
\affiliation{
    \institution{University of California}
    \city{Los Angeles}
    \country{USA}
}

\author{Yufei Ding}
\email{yufeiding@ucsd.edu}
\affiliation{
    \institution{University of California}
    \city{San Diego}
    \country{USA}
}
\begin{abstract}

Modular trapped-ion (TI) architectures offer a scalable quantum computing (QC) platform, with native transport behaviors that closely resemble the Single Instruction Multiple Data (SIMD) paradigm. We present FluxTrap, a SIMD-aware compiler framework that establishes a hardware–software co-design interface for TI systems. FluxTrap introduces a novel abstraction that unifies SIMD-style instructions---inclu-\\ding segmented intra-trap shift SIMD (S3) and global junction transfer SIMD (JT-SIMD) operations---with a SIMD-enriched architectural graph, capturing key features such as transport synchronization, gate-zone locality, and topological constraints. It applies two passes--SIMD aggregation and scheduling---to coordinate grouped ion transport and gate execution within architectural constraints. On NISQ benchmarks, FluxTrap reduces execution time by up to $3.82 \times$ and improves fidelity by several orders of magnitude. It also scales to fault-tolerant workloads under diverse hardware configurations, providing feedback for future TI hardware design.

% Modular trapped-ion (TI) architectures, such as those deployed by Quantinuum, offer high-fidelity qubit transport and scalability via interconnected 1D traps and globally synchronized movement. However, these capabilities introduce unique compilation challenges—including heterogeneous operation latencies, gate-zone locality, and tightly coupled transport constraints—not addressed by existing compiler models. We propose a SIMD-aware compiler framework tailored to these architectures. Our design introduces two key abstractions: segmented intra-trap shift SIMD (S3) and global junction transfer SIMD (JT-SIMD), along with a SIMD-enriched architectural graph that captures gate-zone locality and movement conflicts. Built on these abstractions, we design (1) a SIMD aggregation pass that forms efficient group movements via cost-guided branching, and (2) a SIMD scheduling pass that coordinates inter/intra-trap execution under hardware constraints. Evaluated on NISQ benchmarks, our compiler achieves up to 2.90× speedup and fidelity improvements by several orders of magnitude. We further validate support for FTQC-level circuits under varying hardware configurations.
\end{abstract}
%\begin{CCSXML}
%<ccs2012>
%  <concept_id>00000000.0000000.0000000</concept_id>
%  <concept_desc>Do Not Use This Code, Generate the Correct Terms for Your Paper</concept_desc>
%  <concept_significance>500</concept_significance>
%  %<concept_id>00000000.00000000.00000000</concept_id>
%  <concept_desc>Do Not Use This Code, Generate the Correct Terms for Your Paper</concept_desc>
%  <concept_significance>300</concept_significance>
%  %<concept_id>00000000.00000000.00000000</concept_id>
%  <concept_desc>Do Not Use This Code, Generate the Correct Terms for Your Paper</concept_desc>
%  <concept_significance>100</concept_significance>
 % <concept_id>00000000.00000000.00000000</concept_id>
%  <concept_desc>Do Not Use This Code, Generate the Correct Terms for Your Paper</concept_desc>
%  <concept_significance>100</concept_significance>
%</ccs2012>
%\end{CCSXML}

%\ccsdesc[500]{Do Not Use This Code~Generate the Correct Terms for Your Paper}
%\ccsdesc[300]{Do Not Use This Code~Generate the Correct Terms for Your Paper}
%\ccsdesc{Do Not Use This Code~Generate the Correct Terms for Your Paper}
%\ccsdesc[100]{Do Not Use This Code~Generate the Correct Terms for Your Paper}


\maketitle
\pagestyle{plain}

\section{Introduction}
\label{sect:introduction}
% \begin{figure*}[!t]
% %\begin{figure*}[bp!] [h!]
%         \centering
%         \includegraphics[width=0.98\linewidth]{fig/intro_hardware.pdf}
        
%         \caption{TI architecture combining 1D traps with 2D junctions, with gate zones and auxiliary zones in yellow and gray.}
%         \label{fig:intro_hardware}
% \end{figure*}

Quantum computing (QC) has experienced rapid advancements, with multiple platforms achieving significant breakthroughs in recent years~\cite{arute2019quantum, huang2020superconducting, bravyi2022future, bruzewicz2019trapped, chen2023benchmarking, kok2007linear, bourassa2021blueprint, bluvstein2024logical, wurtz2023aquila}. As each platform offers distinct advantages and continues to advance, competition among them is expected to persist. Among them, trapped ion (TI) systems, which were proposed as one of the original platforms for quantum computing around 30 years ago \cite{cirac1995quantum}, have remained a leading candidate for quantum processors, due to the long coherence time of qubits \cite{03_02_qccd_old_version} and high fidelity of various operations \cite{03_02_new_ion_hardware}.

One of the leading trapped-ion hardware architectures is the \emph{Quantum Charge-Coupled Device (QCCD)} model, which has become the prevailing design for scalable TI systems~\cite{02_25_grid_ion_hardware, 02_26_modular_ion_trap, x_junction0, malinowski2023wire, 03_02_qccd_old_version} and has been adopted by major vendors such as Quantinuum~\cite{03_02_new_ion_hardware, 02_25_grid_ion_hardware}. QCCD architectures feature \textbf{1D ion traps} that support fast ion transport operations, including ion shifts (moving an ion to an empty site) and ion swaps (exchanging the positions of two adjacent ions). These 1D traps are interconnected via \textbf{2D junctions}—cross-shaped structures that enable ion transport across 1D traps—thereby supporting long-range interactions and enabling modular, scalable system layouts.

Unlike superconducting platforms~\cite{superconducting_0, superconducting_1} where qubits are fixed on a 2D grid and interact via static, nearest-neighbor couplings, TI systems utilize qubit transportation for program execution. 
This transport-based architecture enables greater flexibility, but also introduces new physical constraints that compilers must address. In particular: (1) 
% Positions in 1D traps are divided into gates zones and auxiliary zones\cite{02_25_ion_CCD}, with
Gate operations can only be executed when the participating qubits are not only adjacent but also co-located within a designated gate zone along the 1D trap. (2) Different operations (e.g., gates, intra-trap transports, inter-trap transports) have significantly distinct latencies \cite{03_03_ion_operation_time}. 
% (3) Inter-trap transports must be performed globally and cannot be performed with intra-trap transports simultaneously. 
(3) Inter-trap ion transport is globally synchronized \cite{02_25_grid_ion_hardware}: all junctions must perform the same type of transport in the same direction, which cannot conflict with intra-trap operations.

% While these execution rules can be framed as hardware constraints, we view them instead as architectural features that open up new opportunities for optimization. In fact, the transport behavior of QCCD systems—particularly the globally synchronized inter-trap transports—closely mirrors the Single Instruction Multiple Data (SIMD) paradigm in classical computing, where a single instruction governs uniform operations across multiple datapaths~\cite{classical_simd}. Within a single trap, even if the hardware allows flexible individual ion shifts, grouping ion shift operations together (e.g., moving ions in consecutive positions toward an empty slot) can yield better performance by reducing routing depth and supporting efficient pipelining.

While these execution rules can be framed as hardware constraints, we view them instead as architectural features that open up new opportunities for optimization. In fact, the transport behavior of QCCD systems closely mirrors the Single Instruction Multiple Data (SIMD) paradigm in classical computing \cite{classical_simd}. On the inter-trap level, this is reflected in the globally synchronized ion transports, where a single instruction governs uniform operations across all junctions. Within 1D traps with native single-site addressing, the SIMD perspective still offers performance benefits via grouped ion shifts (i.e., moving ions in consecutive positions toward an empty slot), which reduce routing depth and support efficient pipelining. 
% even if flexible individual ion shifts are allowed by the hardware. 
% it can be beneficial to group ion shift operations together (e.g., moving ions in consecutive positions toward an empty slot) due to their shorter latency than intra-trap ion swaps. This grouping can yield better performance by reducing routing depth and supporting efficient pipelining.
% grouping shifts yields better performance and routing efficiency by enabling position reuse and reducing transport overhead. Together, t
These behaviors highlight the need for architectural abstractions and compilation optimizations that are explicitly aligned with the SIMD-style execution on TI systems.
However, existing TI quantum compilers~\cite{02_28_shaper, 03_02_qccdsim} are still fundamentally based on a Single Instruction Single Data (SISD) \cite{SISD} execution model, which lacks the ability to express or leverage the SIMD-style behaviors inherent to QCCD architectures. For example, while these SISD compilers may occasionally produce grouped ion shift patterns, they do not actively search for or exploit them. Even if a SIMD-style grouping can be applied as a post-processing step after initial scheduling, many opportunities for grouped ion transports have already been lost due to early gate-level constraints and suboptimal routing decision. As a result, ion transport instructions often become fragmented and uncoordinated, leading to underutilized transport bandwidth.

In this work, we address two key challenges in compiler design for modular TI systems built on QCCD architectures: (1) how to develop a hardware-compatible abstraction that captures the system’s transport behavior, gate zone constraints, and scheduling exclusivity, and (2) how to design compilation passes that effectively leverage the resulting design space for optimized performance.

To enable a principled hardware--software contract for QCCD systems, we first propose a novel abstraction that reflects their SIMD-style ion transport patterns (Section~\ref{sec: tech1}) consisting of two components:  
% \textcolor{red}{(Jens: I think we can drop ``framework''.)} 
\textit{(1) SIMD-like instruction set.} It models coordinated ion transports both within and across ion trap segments. These include a class of Segmented intra-trap Shift SIMD (S3) instructions, with each instruction dictating grouped \textit{intra-trap} ion shifts in a specified direction, and a class of global Junction Transportation SIMD (JT-SIMD) instructions, with each instruction dictating a global \textit{inter-trap} operation on selected ions along a specified direction. \textit{(2) SIMD-enriched architectural graph abstraction}. It augments the traditional coupling graph by capturing TI execution constraints such as the conflict between intra-trap and inter-trap transports, position-awareness, and thermally constrained gate zones.

\begin{figure*}[t!]
        \centering
        \includegraphics[width=\linewidth, trim=0in 0in 0.2in 0in, clip]{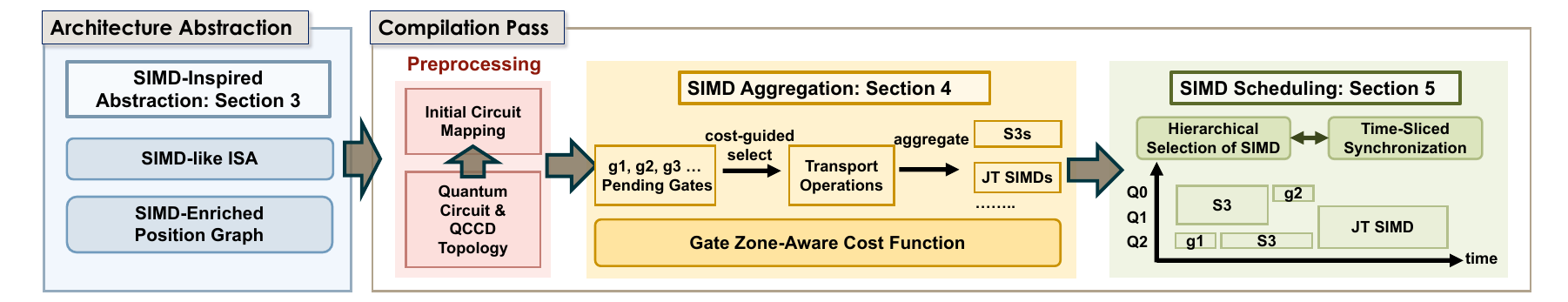}
        
        \caption{Overall Flow of Our Compiler}
        \label{fig:intro_flow}
\end{figure*}

Despite the advantages offered by these abstractions, compile-time optimization under this framework remains highly challenging due to the expansive optimization space and potential conflicts between different SIMD instructions. To this end, we propose a novel TI compiler consisting of two key compilation passes, as shown in Figure~\ref{fig:intro_flow}. 
% \textcolor{red}{(Jens: the text should refer to Figure~\ref{fig:intro_flow}.)} 
The first pass is a \textbf{SIMD Aggregation Pass} (the yellow box in Figure~\ref{fig:intro_flow}) that systematically aggregates scalar ion transport operations into wider SIMD instructions (Section~\ref{sect:simd-aggregation}). With a more global search than greedy approaches guided by a zone-aware cost function, it generates candidate effective SIMD instructions based on the routing needs of ions over the upcoming timesteps. The second pass is a \textbf{SIMD Scheduling Pass} (the green box in Figure~\ref{fig:intro_flow}) that selects and schedules SIMD instructions under hardware constraints (Section~\ref{sect:scheduler}).
At each timestep, it performs a hierarchical decision process that first compares all candidate JT-SIMD instructions and then compares with the cost continuing intra-trap execution, switching to the high-latency inter-trap execution until it becomes urgent or sufficient gain has accumulated.
% , it naturally promotes the grouping of more inter-trap, thus reducing the overall latency of complied programs.
This is coordinated by a time-sliced synchronization mechanism that is aware of the latency disparity between different operations, which enables more efficient scheduling by reducing qubit idleness.

Our contributions in this paper can be summarized as below.
% \vspace{-7pt}
\begin{itemize}
    \item We propose a novel SIMD abstraction that captures the features of modular TI architectures through a set of SIMD-style instructions and a SIMD-enriched architectural graph, along with two compilation passes tailored to TI systems.
    
    \item The SIMD aggregation pass systematically aggregates scalar ion transport operations into wider SIMD instructions via a relatively global search guided by a carefully designed heuristic cost function.
    
    \item The SIMD scheduling pass optimizes the selection and scheduling of SIMD instructions under hardware constraints through hierarchical comparison coordinated by a time-sliced synchronization mechanism.
    
    \item Through comprehensive evaluation, we show that our compiler reduces execution time by up to 3.82 × and improves fidelity by several orders of magnitude on NISQ benchmarks. For FTQC programs, we provide full support and validate its functionality across varied hardware configurations.
    % Through a comprehensive evaluation, we demonstrate that our compiler, incorporating these novel techniques, improves execution time by a factor of 2.90x and significantly enhances fidelity by several orders of magnitude.
    % \textcolor{red}{(Jens: mention that we have both NISQ and FTQC benchmarks.)}
\end{itemize}

\section{Background and Related Work}\label{sect:background}

This section presents essential hardware background and summarizes existing compilation approaches for TI quantum computing. 
%We begin by introducing a 2D grid-based trapped-ion architecture that integrates 1D QCCD traps with 2D transport junctions. We then discuss representative compilation strategies from prior work that reflect varying levels of hardware abstraction.

\subsection{2D Trapped-Ion Hardware Architecture}

% This section presents essential hardware background for modern trapped-ion quantum processors, focusing on the widely adopted QCCD (quantum charge-coupled device) model as implemented in Quantinuum’s latest architecture~\cite{03_02_new_ion_hardware, 02_25_grid_ion_hardware}. This architecture integrates 1D linear ion traps with 2D transport junctions to enable scalable and reconfigurable qubit transport. Our discussion reflects the operational behavior and constraints of real hardware platforms, forming the foundation for the compilation strategies evaluated in this work.
This section provides key hardware background on the QCCD architecture used in modern TI quantum processors\cite{02_25_ion_CCD, 03_02_new_ion_hardware}. The system combines 1D ion traps with 2D junctions to support scalable qubit transport, and introduces specific operational constraints that shape how programs must be compiled and scheduled.

\begin{table}[h]
\caption{Fidelity and Latency Parameters of TI Operations (based on experimental data from~\cite{02_25_ion_CCD, 02_25_grid_ion_hardware, 03_02_new_ion_hardware})}
\resizebox{0.95\linewidth}{!}{
\begin{tabular}{|l||l|l|r|}
\hline
operation types & operation              & fidelity         & latency ($\mu s$)      \\ \hline
\multirow{3}{*}{gate operations} 
& 1Q gate                & 99.9975\%            & 5              \\ 
\cline{2-4} &
2Q gate                & 99.82\%              & 25             \\ 
\cline{2-4} &
measurement              & 99.84\%              & 120             \\ 
\hline 
\multirow{4}{*}{transport operations} &
intra-trap qubit shift & 99.978\%             & 58             \\ 
\cline{2-4} &
intra-trap qubit swap  & 99.978\%             & 200            \\
\cline{2-4} &
inter-trap qubit shift & 99.956\%             & 250            \\ 
\cline{2-4} &
inter-trap qubit swap  & 99.912\%             & 500            \\ \hline
% intra-trap qubit shift & 99.999979\%             & 58             \\ 
% \cline{2-4} &
% intra-trap qubit swap  & 99.999929\%             & 200            \\
% \cline{2-4} &
% inter-trap qubit shift & 99.999911\%             & 250            \\ 
% \cline{2-4} &
% inter-trap qubit swap  & 99.999822\%             & 500            \\ \hline
\end{tabular}
}
\label{tab:hardware_data_table}
\end{table}

\begin{figure}[h!]
        \centering
        \includegraphics[width=\linewidth, trim=0in 0in 0.1in 0in, clip]{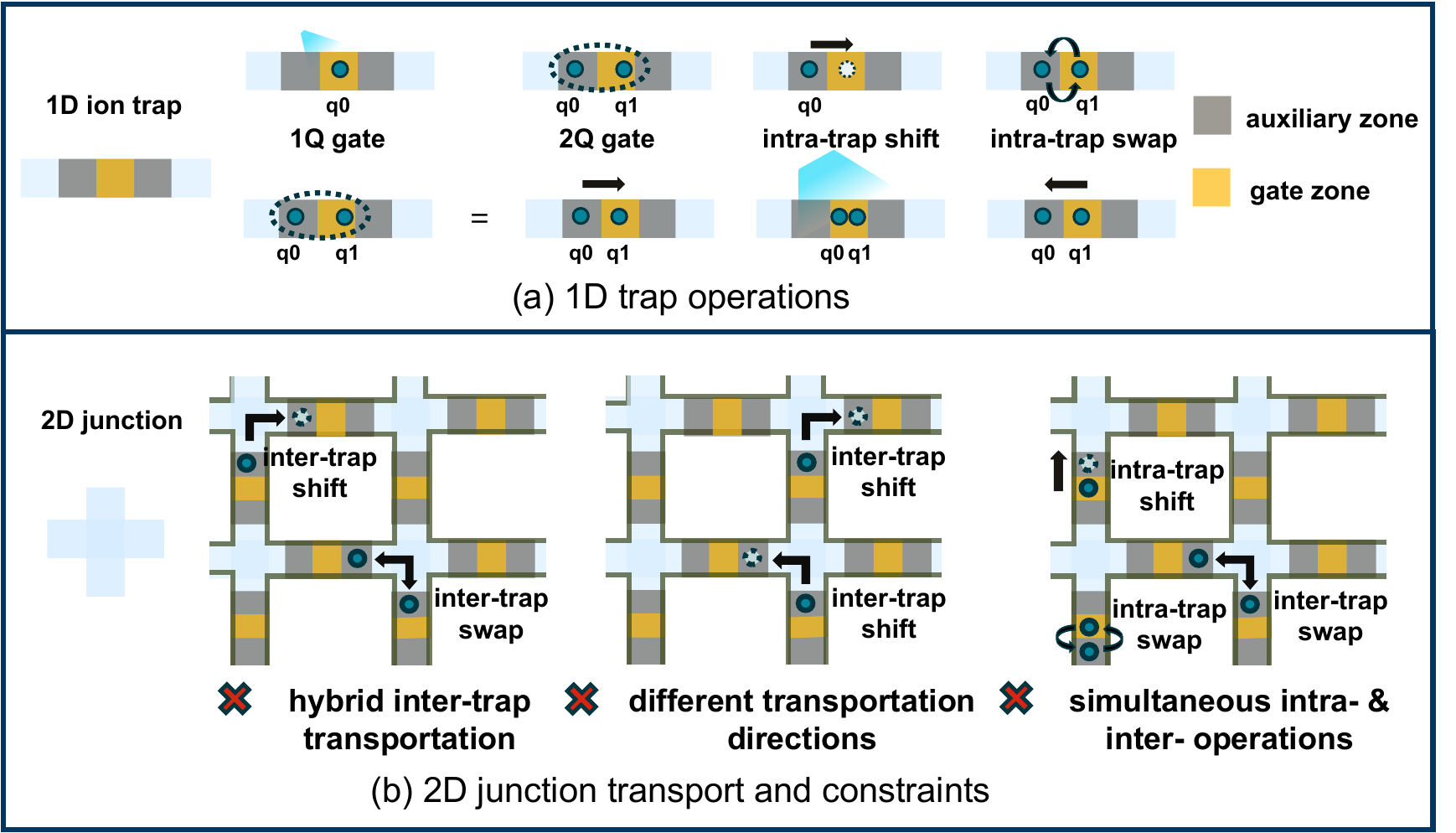}
        
        \caption{Trapped-Ion Transport and Operation Model}
        \label{fig:background}
\end{figure}

\noindent\textbf{1D Linear Traps.}
% \todo{relate gz with thermal}
As illustrated in Figure~\ref{fig:background}(a), each linear trap consists of alternating \textbf{gate zones} (yellow) and \textbf{auxiliary zones} (gray), following a layout widely adopted in Quantinuum’s architecture~\cite{02_25_ion_CCD}.
Gate zones support high-fidelity quantum operations, while auxiliary zones serve as storage locations without gate functionality. The number and placement of gate zones are constrained not only by hardware resources, but also by thermal and optical limitations, as densely packed operations can increase heat dissipation and calibration complexity\cite{gz_thermal}. Therefore, each zone typically holds at most one ion to minimize crosstalk~\cite{03_02_new_ion_hardware}. 

\noindent\textbf{Gate Operations.}
Within a trap, \textbf{gate instructions} include single-qubit (1Q) and two-qubit (2Q) operations, executed via laser pulses~\cite{02_25_ion_CCD} in gate zones. Notably, a 2Q gate only acts on ions co-located in the same gate zone, requiring ions to be shifted in and out for interaction, as shown in Figure~\ref{fig:background}(a). 
% This necessitates \emph{physical ion transport}\cite{03_02_new_ion_hardware}, which introduces greater routing overhead as trap length increases~\cite{02_25_grid_ion_hardware}. 
This implicit transport is accounted for in the gate’s effective latency. As summarized in Table~\ref{tab:hardware_data_table}, 2Q gates generally contribute the most to the overall error due to the complexity of entangling operations, though they remain fast when applied in well-calibrated zones.
% % (Resolved) % \textcolor{red}{(Jens: I suggest that we remove the word ``budget''.)}

\noindent\textbf{Intra-Trap Transport.}
In addition to gate execution, intra-trap transport operations allow repositioning of ions within a trap via two mechanisms: \emph{intra-trap shift}, which moves ions through adjacent empty zones, and \emph{intra-trap swap}, which exchanges positions with a neighboring ion. As shown in Figure~\ref{fig:background}(a) and detailed in Table~\ref{tab:hardware_data_table}, intra-trap shifts are typically faster and more reliable than intra-trap swaps. However, both forms of transport are more latency-intensive than gate execution and should be coalesced or minimized when possible for efficiency.

\noindent\textbf{2D Junction.}
To enable scalable quantum transport beyond isolated 1D traps, modern TI architectures incorporate 2D junctions~\cite{x_junction0,02_25_grid_ion_hardware, x_junction1}, which interconnect linear segments to form a 2D QCCD-style grid~\cite{03_02_new_ion_hardware}. Each 2D junction connects up to four trap segments at orthogonal angles, supporting bidirectional transport in horizontal and vertical directions. 
% This structure generalizes the earlier Y-junction design~\cite{y_junction0, 02_28_shaper, 03_02_qccdsim}, which only allowed T-shaped routing and imposed directional limitations. 
By utilizing such junction-connected segments, the architecture enables modular and dense layout expansion while avoiding long 1D chains. This facilitates reduced transport distances~\cite{02_25_grid_ion_hardware}, increases gate-zone density, and supports more parallel computation.

\noindent\textbf{Inter-Trap Transport.} Inter-trap transport is achieved through two global operations: \emph{(1) inter-trap shift}, where ions move from edge zones into empty positions of neighboring traps (Figure \ref{fig:background}(b)), and \emph{(2) on-site inter-trap swap}, where adjacent edge ions exchange positions (Figure \ref{fig:background}(b)). These operations are globally synchronized: at each cycle, a single transport type (shift or swap) and direction (e.g., left-to-bottom) is selected and broadcast system-wide on all the participating ions at the 2D junctions. 
% Qubits at junction edges may either remain stationary or participate. However, participating qubits must follow the system-wide instruction in both type and direction~\cite{02_25_grid_ion_hardware}.
Compared to intra-trap transport, these inter-trap operations are slower and more error-prone due to longer physical trajectories and increased susceptibility to decoherence, as shown in Table~\ref{tab:hardware_data_table}. 
% Among all transport types, inter-trap swaps typically incur the highest latency and lowest fidelity, followed by inter-trap shifts, while intra-trap shifts remain the most efficient. 

% To manage these costly operations, current QCCD architectures enforce global synchronization constraints on inter-trap transport\cite{02_25_grid_ion_hardware}: only one transport type and direction may be active per cycle, and no intra-trap transport can occur simultaneously. As illustrated in Figure~\ref{fig:background}(b), these design choices simplify hardware control, but place stringent coordination demands on the compiler.
% These physical characteristics motivate compiler strategies that defer costly inter-trap transports when possible, and batch compatible operations to maximize the efficiency of each global transport cycle.
% (Resolved) % \textcolor{red}{(Jens: why is that last sentence here?  I see it as talking about how you got motivated to work on the current paper, but it is in the middle of section on background and related work.  I think you should remove this sentence entirely, or possibly move it to a different section of the paper.)}

In summary, there are three key constraints imposed on inter-trap transport: 
% to simplify system coordination and ensure coherent laser timing: 
(1) only one type of operation (shift or swap) may be issued per cycle; (2) all transport must occur in the same global direction; and (3) intra-trap and inter-trap transport cannot proceed simultaneously due to distinct control pathways, as illustrated in Figure~\ref{fig:background}(b). These constraints places new demands on the software stack, requiring specialized scheduling strategies to coordinate transport efficiently.

\subsection{Previous Compilation Strategies for TIQC}

Many existing approaches~\cite{02_28_shaper,03_02_qccdsim} borrow abstractions or heuristics from superconducting systems~\cite{sabre} and apply them to TI hardware, incorporating some TI-specific constraints but still overlooking key aspects of QCCD architecture. For instance, although each intra-trap transport is lower in cost and more flexible than inter-trap transport, the cumulative volume of intra-trap transport can still be substantial and must be scheduled carefully. Nonetheless, prior compilation strategies have typically overlooked the cost of intra-trap transport, focusing solely on inter-trap routing. 
% (Resolved) % \textcolor{red}{(Jens: I think you should replace ``trivial'' with a different word that is more descriptive and less demeaning.)} 
They have also abstracted away globally synchronized junction usage and sparsely distributed gate zones and instead used simple models that miss major optimization opportunities.
% Further complicating the compilation landscape are architectural constraints such as globally synchronized junction usage~\cite{02_25_grid_ion_hardware} and sparsely distributed gate zones~\cite{02_25_ion_CCD}. These challenges are often abstracted away in simplified models, limiting the applicability of prior solutions to real hardware.

Despite these limitations, prior work introduces useful abstractions and scheduling heuristics for inter-trap routing. We adopt two representative strategies as baselines in our evaluation and adapt them to reflect realistic hardware constraints. One direction, exemplified by \textbf{SHAPER}~\cite{02_28_shaper}, builds on the SABRE compiler—originally developed for superconducting qubits—by modeling ion positions as nodes in a position graph and applying SABRE's swap-based scheduling. While SHAPER focuses primarily on inter-trap routing, we extend its SABRE-based approach to also handle intra-trap transport using a similar swap-to-shift reinterpretation strategy. In contrast, \textbf{QCCDsim}~\cite{03_02_qccdsim} adopts a custom heuristic tailored to trapped-ion inter-trap transport optimizations. It precomputes inter-trap transport paths and performs congestion-aware gate scheduling. We reuse both SHAPER and QCCDsim's mapping and scheduling outputs and augment them with an additional scheduling pass that enforces hardware constraints, including mutual exclusivity between intra- and inter-trap operations and conflict resolution among inter-trap transports with different directions.

\section{SIMD-Inspired Abstractions for Ion Transport} \label{sec: tech1}
% Traditional quantum hardware abstractions—typically consisting of an instruction set architecture (ISA) and a qubit coupling graph—fail to capture many architectural nuances of trapped-ion (TI) machines. In particular, many emerging TI systems, especially those built on QCCD models, exhibit execution behavior that is inherently Single Instruction Multiple Data (SIMD) in nature. Without an abstraction that captures this, current compilation pipelines either underutilize available hardware parallelism or introduce significant control overhead.
% (Resolved) % \textcolor{red}{(Jens: the text in this paragraph is redundant and the points it makes have all been made earlier.  I suggest that you delete this paragraph.  If you notice that something in this paragraph has not been said earlier, then add text to earlier parts of the paper.)}

In this section, we propose a novel abstraction that reflects the SIMD-style ion transport patterns of TI machines. Our abstraction includes: (1) a set of SIMD-like instructions 
% that model coordinated ion transports both within and across ion trap segments,
and (2) a SIMD-enriched architectural graph abstraction. 
% that augments the traditional coupling graph with execution constraints such as dynamically switchable coupling regions, position-awareness, and thermally constrained gate zones. 
This abstraction forms the foundation for the compiler optimizations presented in Section~\ref{sect:simd-aggregation} and Section~\ref{sect:scheduler}.
% (Resolved) % \textcolor{red}{(Jens: what about Section~\ref{sect:scheduler}?)}

\begin{figure*}[]
    \centering
    \includegraphics[width=1\linewidth]{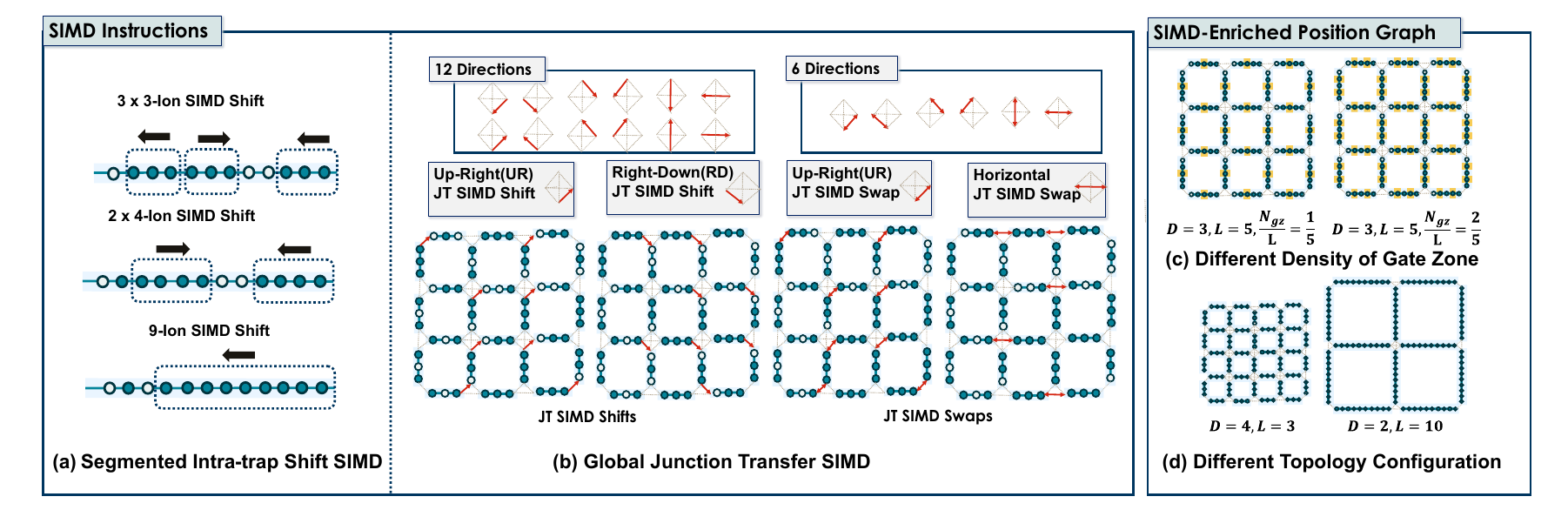}
    \caption{Overview of SIMD Architecture}
    \label{fig:simd}
\end{figure*}

\subsection{SIMD-like ISA}
\label{sec:SIMD}

% but has been treated as scalar by prior compilers, overlooking opportunities for spatial reuse and instruction aggregation. 
% (Resolved) % \textcolor{red}{(Jens: I think Section 3 should avoid bashing the previous compilers; Section 2 did a good job of that, and at some point the bashing gets repetitive.)} 
% We define two classes of SIMD-style instructions—intra-trap and inter-trap movements—as shown in Figure~\ref{fig:simd}.
% The intra-trap SIMD shift, or S3, is natively supported by the 1D trap layout.
% We reinterpret S3 as a true SIMD primitive and leverage it for efficient intra-trap routing.
% Besides, inter-trap JT-SIMD instructions are enabled by a new hardware feature—global 2D control in QCCD architectures—which introduces a new layer of parallelism via coordinated inter-trap transport, explicitly optimized by our compiler.
To enable modular control scheduling and systematic optimization over ion transport, we define a hardware-aligned abstraction layer composed of two SIMD-like instruction classes. These instruction classes serve as ISA-style transport primitives that reflect the physical execution constraints and layout granularity of QCCD architectures (Figure 3):

\vspace{0.3em}\noindent \textbf{1. Segmented Intra-trap Shift SIMD (S3):} 
We define \emph{Segmented Shift SIMD (S3)} as an instruction class that models grouped intra-trap ion shifts. With each linear trap segment acting as a virtual SIMD lane, one or more groups of ions can perform concurrent shifts along various directions. We abstract each grouped shift as an S3 instruction, which specifies: (i) a direction (left or right), and (ii) the ions to shift in this group. The number of ions moved by each S3 instruction is referred to as its \emph{data width}. For example, Figure~\ref{fig:simd}(a) illustrates 3 configurations of grouped ion shifts within a 10-ion segment: three S3 instructions of data width 3, two S3 instructions of data width 4 and one S3 instruction of data width 9.
This instruction class enables compilers to express grouped transport patterns directly, improving intra-trap transport efficiency by aligning software logic with hardware-native execution capabilities.

% \textcolor{red}{(Jens: this entire paragraph of ``Why hardware support this'' seems to fit better in Section 2.)}
% \noindent  \textbf{\textit{Why it matters:}} This abstraction gives software systems a way to express and reason about grouped ion movement. By leveraging S3, the compiler can expose parallelism, reduce routing latency, and balance ion traffic within segments, all while staying aligned with hardware-native capabilities.

\vspace{0.3em}
\noindent \textbf{2. Global Junction Transfer SIMD (JT-SIMD):}  
% Inter-trap ion movement is mediated by junctions in a 2D modular QCCD layout. These junctions allow for ion transfer across trap boundaries, but such transfers are significantly more complex and resource-intensive. We capture these operations using \emph{Junction Transfer SIMD (JT-SIMD)} instructions—each class defining a system-wide transfer mode across junctions (Figure~\ref{fig:simd}(b)).
% \noindent  \textbf{\textit{How it works:}} The hardware supports a total of 18 discrete JT-SIMD classes, covering all allowed ion trajectories through X- and Y-junctions. \todo{explain the three groups and each of the 6 directions within a group}. These are organized into three major categories, shown in Figure~\ref{fig:simd}(b). At each clock cycle, the control system may \emph{activate exactly one} JT-SIMD class across the entire system. While only one transfer class is enabled at a time, each junction or ion group may locally choose to participate or remain idle. This results in a \emph{variadic} SIMD instruction: the operation type is globally fixed, but its spatial footprint is circuit-dependent.
We define \emph{Junction Transfer SIMD (JT-SIMD)} as an instruction class that models globally synchronized inter-trap ion transports. Each JT-SIMD instruction encodes a system-wide transport mode across all 2D junctions in a QCCD grid, which specifies: (i) a transport type (shift or swap), (ii) a global transport direction (e.g., left-to-right, top-to-left) and (iii) the junctions that participate the transport (each individual junction may choose to participate or remain idle). According to (i) and (ii), JT-SIMD instructions are categorized into 18 subclasses---12 for directional shifts and 6 for directional swaps---each corresponding to a legal routing path through the X-junctions (Figure~\ref{fig:simd}(b)). At each timestep, only one JT-SIMD instruction can be issued, reflecting the hardware-level constraint of global synchronization. This instruction class enables the compiler to reason about long-range transport with explicit awareness of global coordination requirements.

\subsection{SIMD-Enriched Position Graph}

% Our instruction-level abstractions are accompanied by a refined architectural graph model that generalizes and extends classical coupling graphs (commonly used in superconducting systems)~\cite{Qiskit} and recent position-graph models proposed for ion-trap architectures~\cite{02_28_shaper}.
% %
% In a \emph{position graph}, nodes represent physical locations rather than logical qubits, allowing for dynamic occupancy by ions. This enables a key distinction between \emph{shift} operations (movement into an empty site) and \emph{swap} operations (movement between two occupied sites). While insightful, this representation is still incomplete in capturing the full hardware semantics. We introduce three major extensions to develop a SIMD-enriched coupling graph abstraction.
% Our instruction-level abstractions are paired with a refined architectural graph model that extends classical coupling graphs (used in superconducting systems~\cite{Qiskit}) and position-graph models (for TI architectures~\cite{02_28_shaper}). In a \emph{position graph}, nodes represent physical locations (not logical qubits), enabling dynamic ion occupancy and distinguishing \emph{shift} (into empty sites) from \emph{swap} (between occupied sites) operations. While this model is useful, it remains incomplete in capturing full hardware semantics. Here, we introduce three key extensions to build a SIMD-enriched coupling graph abstraction.
Our instruction-level abstractions are paired with a refined architectural graph model that extends previous \emph{position graph} models for TI architectures~\cite{02_28_shaper}. In a position graph, nodes represent physical locations in the 1D traps, rather than logical qubits as in coupling graphs widely used in superconducting systems~\cite{Qiskit}. This reflects the dynamic ion occupancy in TI systems and naturally distinguishes \emph{shift} (into empty sites) from \emph{swap} (between occupied sites) operations. While this model is useful, it remains incomplete in capturing full hardware semantics. Here, we introduce three key extensions to build a SIMD-enriched coupling graph abstraction.

\vspace{0.3em}
\noindent  \textbf{1. Dynamic Switchable Position Graph:} The conflict between different JT-SIMD instructions and that between JT-SIMD and intra-trap operations require a dynamic notion of connectivity. Rather than using a static position graph, we define a \emph{switchable position graph} that changes dynamically with the active instructions. At any given time step, the position graph switches between an intra-trap mode which only consists of intra-trap couplings and an inter-trap mode which only consists of inter-trap couplings in a specified direction. 
This dynamic coupling model is similar to reconfigurable interconnect fabrics in classical architectures and more accurately reflects the time-varying transport capabilities of QCCD systems. It avoids inefficient routing by ensuring a natural compliance with the hardware constraints.

% The presence of global JT-SIMD operations requires a dynamic notion of connectivity. Rather than using a static position graph, we define a \emph{switchable position graph} that changes at runtime based on the active JT-SIMD instruction. At any given time step, only the transport patterns allowed by the current JT-SIMD class are considered legal paths. This dynamic coupling model is similar to reconfigurable interconnect fabrics in classical architectures and more accurately reflects the time-varying transport capabilities of QCCD systems. It avoids the inefficiencies of speculative routing and enables the compiler to reason about time-aware connectivity.

\vspace{0.3em}
\noindent  \textbf{2. Gate Zone integrated Position Graph:} In traditional coupling graphs for superconducting devices, each node represents a qubit and each edge indicates the ability to apply a two-qubit gate between connected qubits. %In superconducting systems, connectivity implies the ability to perform both movement and gate operations.
However, in ion-trap systems, these capabilities are decoupled. While ions may move along an edge in a position graph, gate operations can only occur within designated \emph{gate zones}. This distinction is yet missed in previous abstraction~\cite{02_28_shaper}. Figure~\ref{fig:simd}(c) illustrates different gate zone placements in various densities. In large-scale hardware, gate zones are typically placed sparsely along linear trap segments to reduce control, calibration and thermal overhead~\cite{thermal_noise_duan,gz_thermal}. 
Our abstraction explicitly encodes gate zone locations within the position graph, ensuring the compiler's awareness of these dedicated locations. More importantly, it enables a feedback loop between software and hardware: software-level scheduling can alleviate gate zone congestion, and hardware teams can adapt gate zone layout based on compiler-informed guidance---a co-design opportunity not captured by previous abstractions.

% \begin{figure*}[!h]
%     \centering
%     \includegraphics[width=1\linewidth]{fig/position_graph.pdf}
%     \caption{}
%     \label{fig:position_graph}
% \end{figure*}

\vspace{0.3em}
\noindent  \textbf{3. Configurable Trap and Junction Topology:} Beyond logical constraints, our abstraction enables software visibility into key physical design choices---specifically, the tradeoff between extending linear trap segments versus increasing the number of 2D junctions. As illustrated in Figure~\ref{fig:simd}(d), multiple architectures may host the same number of ions using different topological configurations: longer intra-trap segments with fewer junctions, or more frequent junctions with shorter trap segments. These decisions influence hardware complexity, control parallelism, and thermal scalability. By exposing these design axes explicitly, our abstraction allows software and compiler strategies to target hardware-specific tradeoffs, enabling layout-sensitive scheduling, and traffic-aware routing.

\section{SIMD Aggregation Pass}\label{sect:simd-aggregation}

% With SIMD-style instructions and enriched architectural abstractions in place, the compiler must translate quantum programs into hardware-efficient execution plans that expose parallelism, respect architectural constraints, and minimize routing cost. The first step in this process is to aggregate scalar ion movement operations into SIMD-style instructions. Rather than issuing ion transports one-by-one, we analyze spatial and temporal movement patterns to detect opportunities for concurrent execution. This section details 
% % how we form intra-trap S3 and inter-trap JT-SIMD instructions via a staged aggregation strategy 
% how intra-trap S3 and inter-trap JT-SIMD operations are formed by analyzing routing needs of upcoming gates with a gate-zone-aware cost function.

With SIMD-style instructions and architectural abstractions defined, the compiler must generate execution plans that expose parallelism, respect hardware constraints, and reduce routing cost. The first step is to aggregate scalar ion transports into SIMD-style instructions by analyzing spatial patterns and upcoming gate demands. This section presents how intra-trap S3 and inter-trap JT-SIMD operations are formed using a gate-zone-aware cost function.

\begin{figure}[]
        \centering
        \includegraphics[width=\linewidth, trim=0.1in 0.4in 0.3in 0.2in, clip]{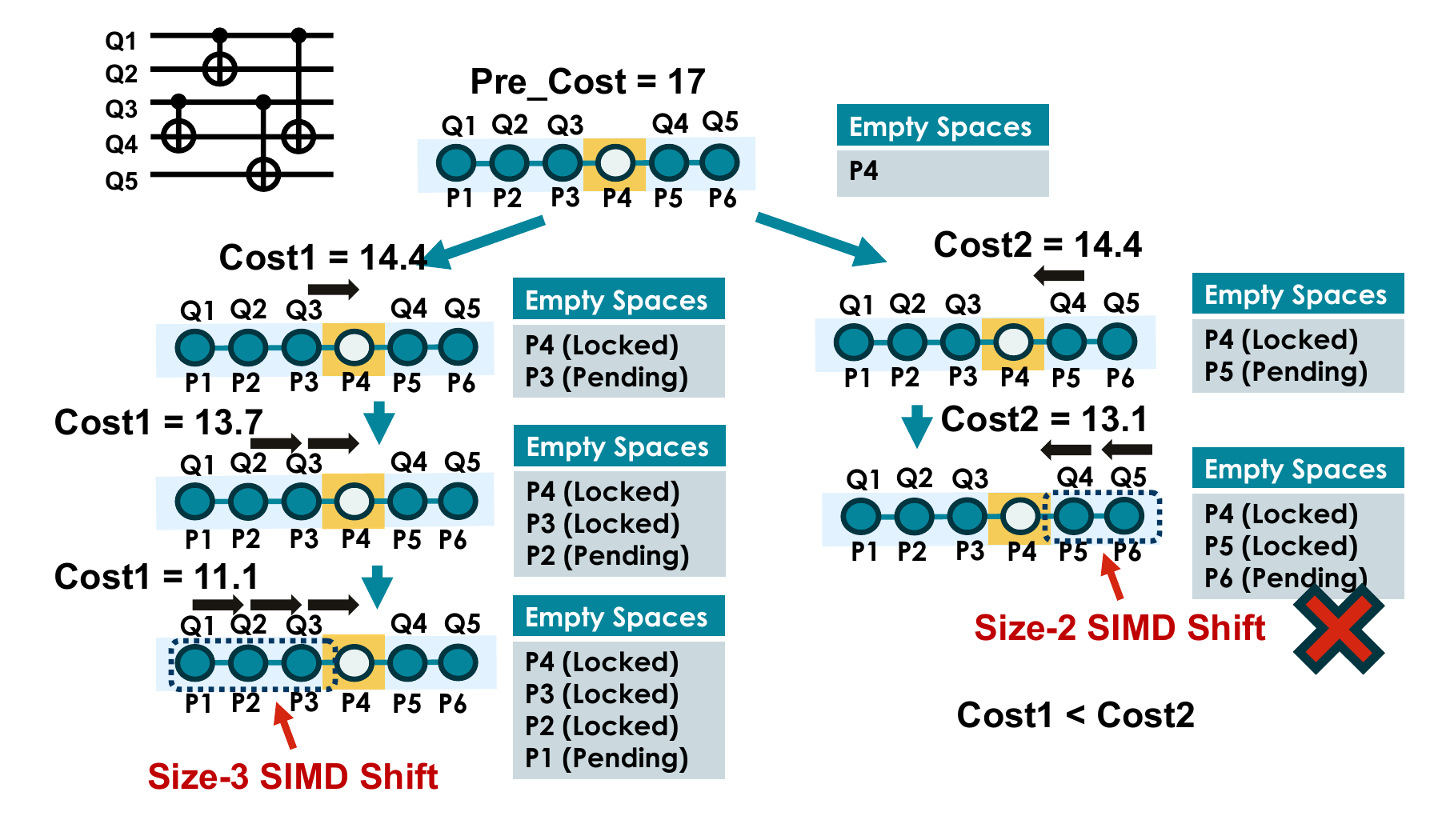}
        
        \caption{SIMD-Aware S3 Formation with Position Reuse}
        \label{fig:group_shift}
\end{figure}

\subsection{Instruction Aggregation Strategies}
% To exploit SIMD-style transport, the compiler identifies scalar ion movements that can be aggregated into wide parallel instructions. This section outlines how intra-trap S3 and inter-trap JT-SIMD operations are formed by analyzing routing needs of upcoming gates, enabling coordinated and efficient movement.

\noindent \textbf{1. Intra-Trap S3 Aggregation:}
To maximize intra-trap transport efficiency, our approach proactively constructs shift sequences that leverage position reuse—each selected shift frees its source position for subsequent moves. The process begins by scanning pending gates to identify qubits whose movement in a particular direction would reduce their heuristic cost. For example, in Figure~\ref{fig:group_shift}, both Q3 and Q4 intend to shift into the same empty position, leading to a potential conflict.
Rather than resolving this greedily, we initiate a timeline-aware forward search to explore viable shift sequences. Starting with Q3, we trace a path where Q2 and Q1 can shift ahead of it in the same direction. As each shift is selected, we lock the destination position to prevent further contention and release the source position to allow follow-up shifts. This forms a candidate group shift {Q1, Q2, Q3}. A similar process applied to Q4 reveals another candidate group shift {Q4, Q5}.
In general, this process produces multiple group shifts across the 1D trap. Some groups are independent—they operate in different regions and do not compete for the same empty space—and can thus be scheduled in parallel. Others, like {Q1, Q2, Q3} and {Q4, Q5}, conflict due to shared dependencies, and must be evaluated against one another. We use our gate zone-aware heuristic to compare the total benefit of each conflicting group, and choose the one that provides the greatest forward progress. In this case, {Q1, Q2, Q3} achieves a lower heuristic cost and is selected for execution.

\noindent \textbf{2. Inter-Trap JT-SIMD Aggregation:}
We examine inter-segment routing requirements and generate instructions from the 18 predefined JT-SIMD classes. For each possible global transport (e.g., down shift, upright shift), we evaluate qubits involved in upcoming two-qubit gates and determine whether shifting or remaining stationary yields a lower heuristic cost. The ions selected for transport under each operation are then aggregated into a single JT-SIMD instruction. For example, in the case of a global down shift, we compare the cost of shifting each qubit downward versus keeping it in place, and aggregate those benefiting from downward movement into a unified instruction.

\subsection{Gate Zone-Aware Cost Function}

\begin{figure}[h!]
        \centering
        \includegraphics[width=0.85\linewidth, trim=0.1in 0.3in 0.3in 0.1in, clip]{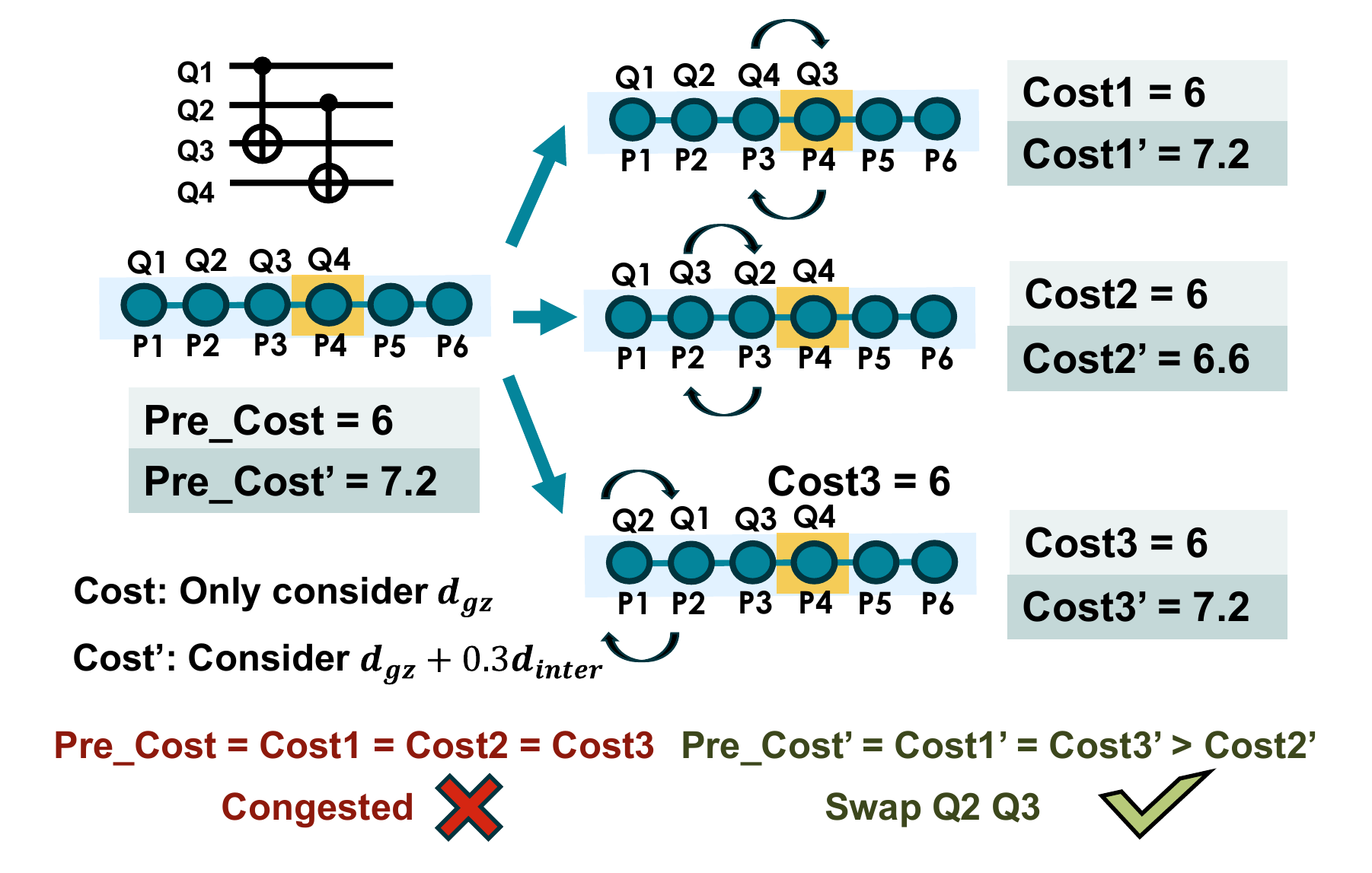}
        
        \caption{ Alleviating Gate Zone Congestion via Spacing-Aware Prioritization}
        \label{fig:congestion}
\end{figure}

To effectively guide SIMD aggregation and transport scheduling, our compiler employs a gate zone-aware heuristic that integrates both spatial and temporal priorities. This heuristic prioritizes gates whose qubits are positioned favorably with respect to their assigned gate zones and penalizes those with higher routing overhead. Specifically, it evaluates each gate in the front layer based on (1) the cumulative distance from participating qubits to the target gate zone, $d_{\text{gz}}(g)$, and (2) the relative distance between qubits for two-qubit gates, $d_{\text{inter}}(g)$, to discourage zone congestion, as shown in Figure~\ref{fig:congestion}. The second term is weighted by a coefficient $\alpha$ and included only when the gate is a two-qubit operation. This selective inclusion is controlled by the indicator $\mathbb{I}_{\text{tg}}(g)$, which is set to 1 only for two-qubit gates.
We set $\alpha = 0.3$ in our implementation to reflect that inter-qubit distance is not the dominant factor during placement and routing. This ensures that the compiler prioritizes minimizing gate zone distance while still incorporating spacing when congestion arises.
To reduce search overhead, each pending gate is assigned a target zone only once during compilation. Reassignment of gate zones is triggered only under excessive routing delays, preserving scalability. To further mitigate congestion, gates targeting overloaded zones are deprioritized, and in cases of persistent congestion, our scheduler may temporarily ignore some gates that require future transport assistance when computing the heuristic cost.
Formally, the heuristic cost is given by:
\begin{equation}
\mathcal{C(G, \pi)} \;\;=\;\; \sum_{g \in F, \ g.\text{qubits} \notin E} \left( d_{\text{gz}}(g) + \alpha \cdot \mathbb{I}_{\text{tg}}(g) \cdot d_{\text{inter}}(g) \right)
\end{equation}
% (Resolved) % \textcolor{red}{(Jens: in a quantum paper, $H$ is a dicy name :-); I think a better name is $C$, for cost function.)}

Here, $G$ denotes the SIMD-enriched position graph that encodes zone-level connectivity and constraints, and $\pi$ is the current mapping from logical qubits to physical positions. $F$ denotes the set of ready gates in the dependency graph (the front layer), and $E$ is the set of qubits currently involved in in-flight operations. 
Unlike traditional compilers—which overlook gate zone availability and rely on sparse ion placement to avoid contention—our heuristic enables high-density layouts and efficient gate scheduling by dynamically co-optimizing qubit transport and gate placement. It integrates trap-level congestion and transport-induced scheduling delays into a unified cost model, serving as a simulator-driven feedback loop during compilation.

\section{SIMD Scheduling Pass}
%\section{SIMD Scheduling Strategies}
\label{sect:scheduler}

% To execute aggregated SIMD instructions efficiently, the compiler must schedule them under hardware constraints such as control exclusivity and thermal limitations \cite{gz_thermal}.

% \textcolor{red}{(Jens: here the paper mentions the thermal limitations yet again, but so far without giving much detail about what those limitations really are.)} 
% This section presents a two-stage strategy: first, a hierarchical decision process selects the optimal JT SIMD instruction class and determines whether to prioritize intra- or inter-trap operations; second, a time-sliced SIMD synchronization mechanism coordinates operations at the sub-cycle level, improving latency handling and overall throughput. Together, these scheduling techniques maximize gate readiness while minimizing system contention and routing delay.

To efficiently execute aggregated SIMD instructions, the compiler must schedule them under hardware constraints such as control exclusivity and thermal limits~\cite{gz_thermal}, which stem from limited laser power and gate zone heating.
This section introduces a two-stage scheduling strategy: (1) a hierarchical decision process that selects the best JT-SIMD instruction and determines whether to continue intra-trap or switch to inter-trap transport, and (2) a time-sliced synchronization mechanism that coordinates execution at a fine-grained level to improve latency management and throughput.
Together, these techniques enhance gate readiness while reducing contention and routing delays.

\subsection{Hierarchical Selection of SIMD Instructions}

This subsection outlines a hierarchical scheduling strategy that governs which SIMD instruction types to issue at each timestep. It first selects the optimal JT-SIMD class based on projected layout improvement, then dynamically decides between intra-trap and inter-trap instructions by balancing latency, thermal cost, and global routing benefit.

\noindent \textbf{1. Selection of JT-SIMD Class per Cycle:} At each clock cycle, only one JT-SIMD class is allowed. Our scheduler calculates the heuristic score of the 18 pre-defined JT-SIMD instructions, with a lookahead into operations over the next few timesteps. Then it schedules the one with the lowest cost, enabling the most progress in terms of ready-to-route ions.

% Our scheduler uses a lookahead window to evaluate which JT-SIMD class enables the most progress in terms of ready-to-route ions. A priority queue ranks classes by aggregate ion distance reduction and alignment with future gates.

% we examine the 18 pre-defined JT-SIMD instructions and choose the one with the lowest cost according to our heuristic function.
\begin{figure*}[]
        \centering
        \includegraphics[width=\linewidth, trim=0in 0.2in 0.1in 0in, clip]{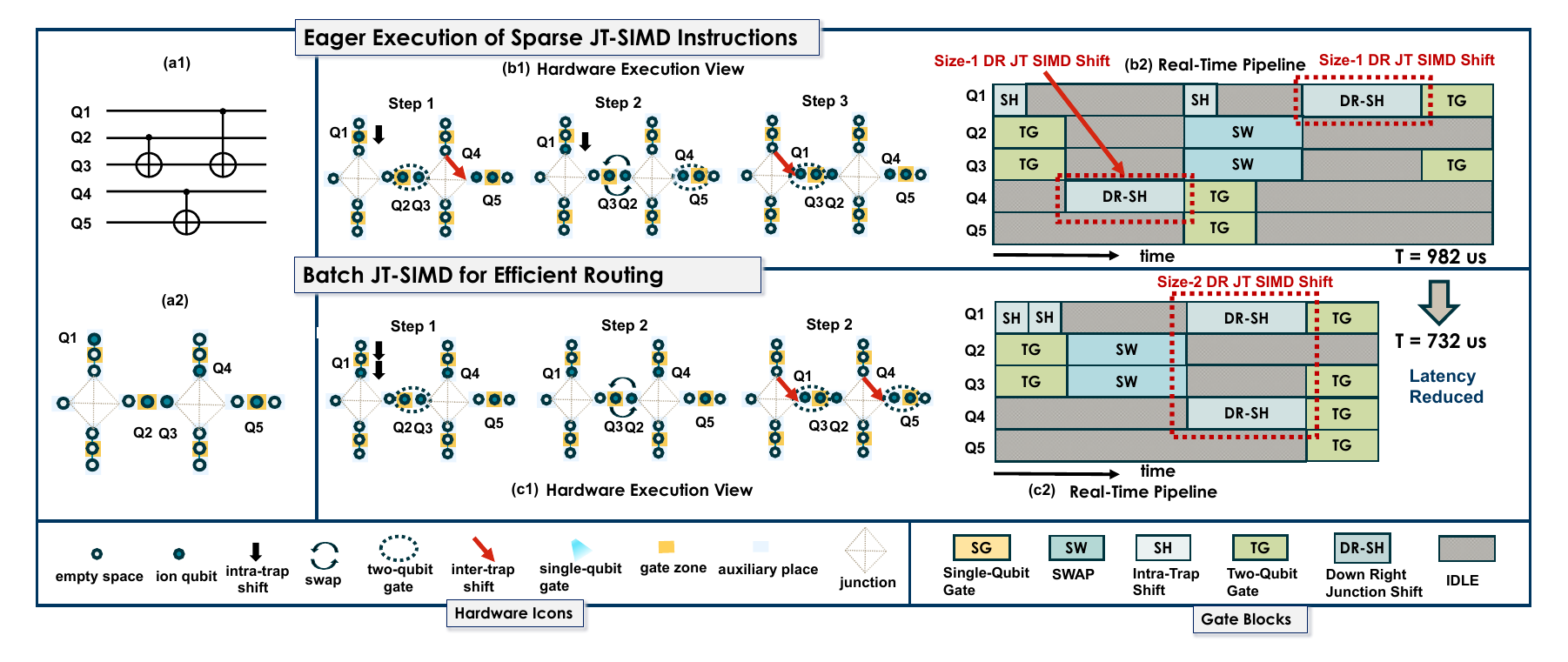}
        
        \caption{
        Better SIMD Scheduling via Optimized Intra–Inter Transport Switching: This example illustrates the benefit of our compiler’s decision rule, which activates JT-SIMD instructions only when their routing gain exceeds that of intra-trap alternatives, enabling more effective batching. The circuit and its initial layout are shown in (a1) and (a2). (b1) and (b2) depict the baseline scheduler, which greedily triggers inter-trap movements, resulting in fragmented transport. In contrast, our optimized scheduler in (c1) and (c2) delays and aggregates JT-SIMD instructions to align transport directions, improving SIMD efficiency and reducing overhead.
        % (Resolved) % \textcolor{red}{(Jens: Figure 6(c2) says $682\mu s$; I wonder which one of 662 and 682 is correct.)}
        }

        \label{fig:jt_aggregation}
\end{figure*}

\noindent \textbf{2. Intra-Trap versus Inter-Trap Prioritization:} In trapped-ion systems, intra-trap and inter-trap SIMD instructions cannot be executed simultaneously due to hardware-level control restrictions. Moreover, inter-trap transport operations typically incur 2--3$\times$ higher latency and larger thermal overhead compared to intra-trap shifts. To effectively coordinate these transport types, our scheduler dynamically selects between intra-trap S3 and inter-trap JT-SIMD instructions by comparing their potential to reduce the overall heuristic cost.
At each scheduling step, the compiler evaluates two candidate transport plans: one composed purely of intra-trap S3 shifts, and another that includes inter-trap JT-SIMD instructions. The scheduler prioritizes intra-trap S3 instructions for their lower latency and less thermal overhead, while activating inter-trap JT-SIMD opportunistically to unlock critical inter-trap routing progress. Specifically, global inter-trap transport is initiated only when its benefit to routing significantly outweighs that of intra-trap alternatives. Let $\mathcal{C}(G, \pi)$, as mentioned in Section \ref{sect:simd-aggregation}, denote the current heuristic cost of the system layout, and let $\pi_{\text{intra}}$ and $\pi_{\text{inter}}$ represent the resulting qubit mappings after applying intra-trap and inter-trap transports, respectively. 
% \textcolor{red}{(Jens: this is the same $\mathcal{C}$ as in Equation (1), so Equation (1) should define $\mathcal{C}$ with the same two parameters as here.)} 
The condition of selecting inter-trap JT-SIMD can be expressed as:
\begin{equation}
\mathcal{C}(G, \pi) - \mathcal{C}(G, \pi_{\text{inter}}) > 2 \cdot (\mathcal{C}(G, \pi) - \mathcal{C}(G, \pi_{\text{intra}}))
\label{eq:decision-rule}
\end{equation}
This decision rule naturally enables the aggregation of more inter-trap transports, thereby enhancing the transport efficiency.
Fig.~\ref{fig:jt_aggregation} demonstrates this benefit with an example of implementing the circuit in Fig.~\ref{fig:jt_aggregation}(a1) with an initial layout in Fig.~\ref{fig:jt_aggregation}(a2). 
A traditional compiler has no decision rule like Rule~(\ref{eq:decision-rule}) and would eagerly execute JT-SIMD instructions once their cost is lower than S3 instructions.
% Without this decision rule, compilers may eagerly execute JT-SIMD instructions once their cost is lower than S3 instructions. 
This would lead to an operation scheduling with the steps shown in Fig.~\ref{fig:jt_aggregation}(b1).
\textbf{Step 1:} the 2Q gate between Q2 and Q3 is first conducted as these two qubits are adjacent. Meanwhile, Q1 and Q3 can be brought closer by an intra-trap down shift on Q1, while Q4 and Q5 can be brought closer by an inter-down-right shift on Q4. Due to hardware constraints, the latter one is scheduled after the former one is complete, as shown in Fig.~\ref{fig:jt_aggregation}(b2). \textbf{Step 2:} The 2Q gate between Q4 and Q5 is conducted since the two qubits are adjacent now. Concurrently, Q1 and Q3 are brought further closer by further shifting down Q1 and 
swapping Q2 and Q3. \textbf{Step 3:} An inter-trap shift on Q1 is conducted to bring Q1 adjacent to Q3, so that the 2Q gate between Q1 and Q3 can be conducted now.
Taking the latency of each operation into account, this scheduling can be translated to that in Fig.~\ref{fig:jt_aggregation}(b2), leading to an overall computation time of $982\mu s$. 
% (Resolved) % \textcolor{red}{(Jens: Figure 6(b2) says $882\mu s$; I wonder which one of 862 and 882 is correct.)} 
Note that while the two inter-trap shifts are in the same direction, they are conducted separately due to the eager execution of JT-SIMD instructions.

In contrast, with the decision rule above, our compiler can provide an optimized scheduling that identifies and groups inter-trap transports along the same direction. With the same example, our compiler would schedule operations in the following steps, as represented by Fig.~\ref{fig:jt_aggregation}(c1). \textbf{Step~1:} The 2Q gate between Q2 and Q3 is first conducted as these qubits are adjacent. Concurrently, Q1 is shifted down for two steps to get closer to Q3. \textbf{Step 2:} Q2 and Q3 are swapped to bring Q3 closer to Q1. \textbf{Step 3:} Q1 and Q4 are shifted in the down-right direction by a global inter-trap transportation to bring them adjacent to Q3 and Q5, so that the two 2Q gates between Q1,Q3 and between Q4,Q5 can be conducted.
The primary optimization in these steps is that our compiler delays and groups the inter-trap shifts on Q1 and Q4 to reduce transportation latency. As shown in Fig.~\ref{fig:jt_aggregation}(c2), this reduces the overall computation time to $732\mu s$ only.
% (Resolved) % \textcolor{red}{(Jens: Figure 6(c2) says $682\mu s$; I wonder which one of 662 and 682 is correct.)}

\subsection{Time-sliced SIMD Synchronization}

\noindent In a scheduling process, operations may complete at different times due to their distinct latencies, effectively dividing execution into fine-grained time slices. To improve scheduling within each time slice, we propose a synchronization mechanism that enhances scheduling precision and minimizes qubit idle time by tracking real-time execution information.

This mechanism handles three types of information at each time slice: (1) qubit usage — it tracks which qubits are engaged in operations and which are idle, preventing the scheduling of operations that require the currently engaged qubits; (2) position availability — it monitors which positions are occupied or free, enabling immediate scheduling of operations once space is freed by other operations; and (3) program dependencies — it maintains a dynamic dependency graph for the program to identify executable operations as soon as their prerequisites complete.

To determine how time advances within the scheduling cycle, we adopt different strategies depending on the type of chosen transport. For intra-trap transport, we use the shortest remaining time (SRT) approach, where the scheduler advances time by the duration of the operation that finishes earliest. This allows the system to quickly release qubits and space resources, enabling other operations to proceed without unnecessary delay. In contrast, when inter-trap transport is scheduled, the system switches to a longest remaining time (LRT) strategy, waiting for all in-flight intra-trap operations to complete before initiating any inter-trap transport. This avoids execution conflicts due to hardware constraints and allows inter-trap operations to be executed in a clean, batched manner.
\begin{figure}[]
        \centering
        \includegraphics[width=\linewidth, trim=0.25in 0.3in 0.25in 0.2in, clip]{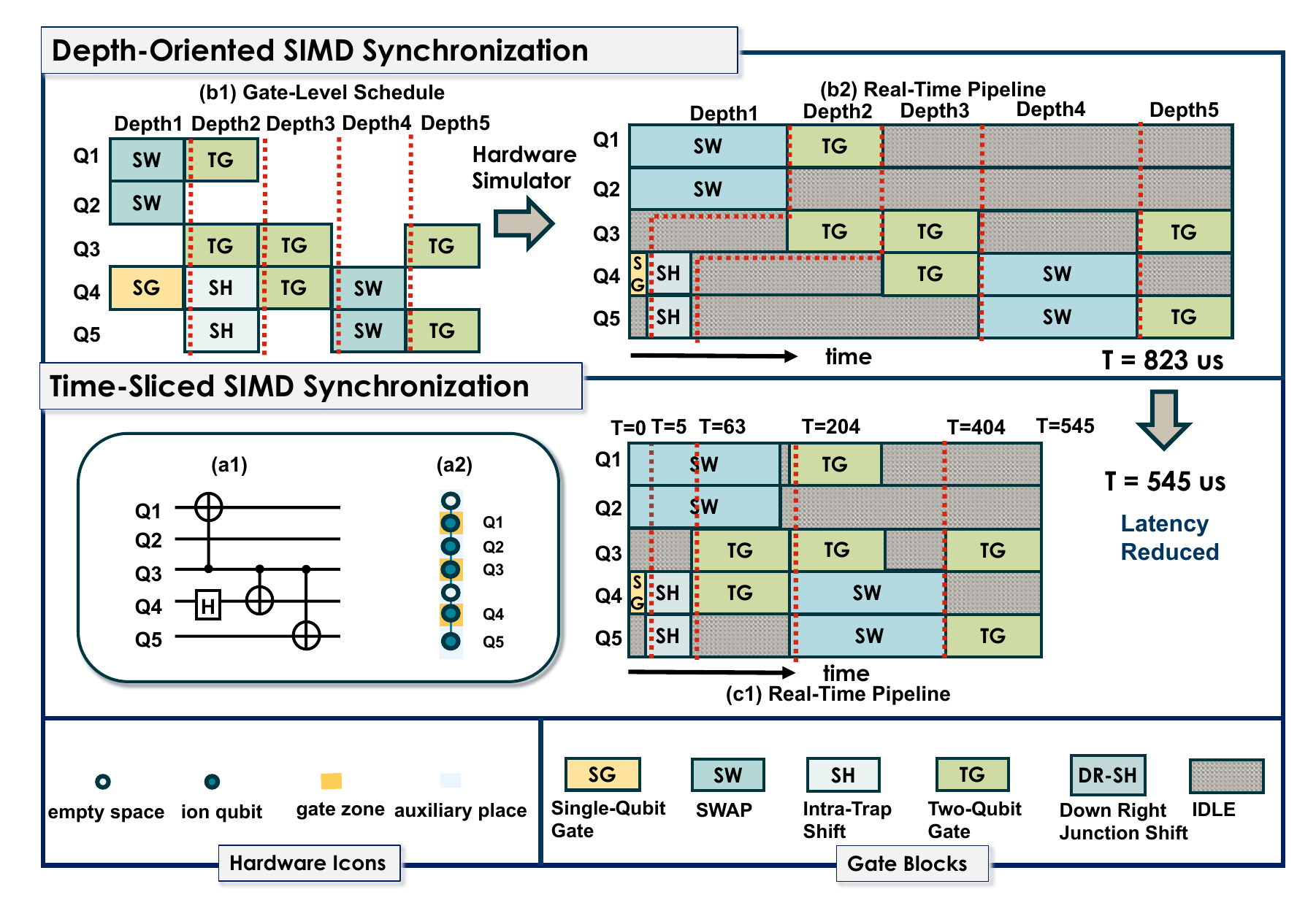}
        
        \caption{Improved Execution via Time-Sliced SIMD Synchronization: This example demonstrates how our fine-grained scheduler overlaps operations by leveraging real-time context. (a1) and (a2) show the circuit and initial layout. The baseline plan (b1, b2) follows depth-based scheduling and leaves idle gaps.
        % (Resolved) % \textcolor{red}{(Jens: Figure 7(c1) says $545\mu s$; I wonder which one of 525 and 545 is correct.)}
        }

        \label{fig:time_sync}
\end{figure}

Together, these mechanisms provide the scheduler with up-to-date execution context and enable precise control over transport execution, improving both temporal utilization and overall circuit efficiency.
The advantage of our fine-grained synchronization can be illustrated by an example in Fig.~\ref{fig:time_sync}, where the TI systems implements the circuit in Fig.~\ref{fig:time_sync}(a1) with an initial layout in Fig.~\ref{fig:time_sync}(a2). 
A traditional compiler has no awareness of latency disparity and would employ depth-oriented synchronization, with operations scheduled in the steps shown in Fig.~\ref{fig:time_sync}(b1). 
In the traditional strategy (Fig.\ref{fig:time_sync}(b1)), the compiler assumes all operations in the same step complete simultaneously, thus missing opportunities for overlap. 
\textbf{Step 1:} The 1Q gate is conducted on Q4. Concurrently, Q1 and Q3 are brought adjacent by swapping Q1 with Q2. \textbf{Step 2:} 
% The 2Q gate between Q1 and Q3 is conducted. Meanwhile, Q4 is shifted up to become adjacent to Q3 and Q5 is shifted up to get closer to Q3.
With the 1Q gate and the Q1–Q2 swap presumed completed, the compiler proceeds to schedule the 2Q gate between Q1 and Q3. At the same time, Q4 and Q5 are shifted upward in preparation for subsequent gates with Q3.
\textbf{Step 3:} The 2Q gate between Q3 and Q4 is conducted. \textbf{Step 4:} Q5 is brought adjacent to Q3 by swapping with Q4. \textbf{Step 5:} The 2Q gate between Q3 and Q5 is conducted. Taking the latency of each operation into account, this scheduling can be translated to that in Fig.~\ref{fig:time_sync}(b2), leading to an overall computation time of $823\mu s$.

In contrast, with the time-sliced synchronization, our compiler can produce a better scheduling as shown in Figure~\ref{fig:time_sync}(c1), which has the following steps. \textbf{Step 1:} The 1Q gate is conducted on Q4, while Q1 and Q3 are brought adjacent by swapping Q1 with Q2. \textbf{Step 2:} 
% A grouped shift is conducted on Q4 and Q5 in the up direction to bring them closer to Q3.
Notably, our compiler detects that the 1Q gate on Q4 has already finished while the swap is still ongoing, and immediately schedules a grouped S3 shift on Q4 and Q5 to move them toward Q3—an opportunity missed by depth-based synchronization.
% \textcolor{red}{(Jens: the description is good, but why is the text different from the text about the same thing in Figure~\ref{fig:time_sync}(b2)?  I would prefer that those two pieces of text become more similar, to avoid confusing the reader who will see the same depiction of the shifts in Figure~\ref{fig:time_sync}(b2) and c1).)} 
\textbf{Step 3:} The 2Q
gate between Q3 and Q4 is conducted, considering that the swap gate between Q1 and Q2 is still executing. \textbf{Step 4:} After the 2Q gate between Q3 and Q4 is completed, we then swap Q4 and Q5 to bring Q4 adjacent to Q3, and conduct the 2Q gate between Q1 and Q3. \textbf{Step 5:} This is followed by the 2Q gate between Q3 and Q5. 

The main difference between our scheduling and the depth-oriented scheduling is that we fill up the time of the swap operation between Q1 and Q2 with the execution of 2Q gate between Q3 and Q4. We remark that this cannot be identified by compilers with a depth-oriented synchronization, despite that the 2Q gate between Q1 and Q3 and that between Q3 and Q4 are actually commutable. This is because as shown in \ref{fig:time_sync}(b1), they would think that the shift on Q4 has to be initiated after the completion of swap between Q1 and Q2, hence the 2Q gate cannot between Q3 and Q4 are not executable until the next depth. This forces the compiler to choose the 2Q gate between Q1 and Q3 in the current depth. Overall, our time-sliced synchronization leads to a shorter computation latency of $545\mu s$.
% (Resolved) % \textcolor{red}{(Jens: Figure 7(c1) says $545\mu s$; I wonder which one of 525 and 545 is correct.)}

% \input{05_tech-2}
% \input{05_tech}
\section{Evaluation}\label{sect:eval}

% This section presents a comprehensive evaluation of our SIMD-aware compiler, highlighting its ability to co-optimize compilation and hardware execution for trapped-ion quantum systems.
This section evaluates our SIMD-aware compiler and demonstrates its effectiveness in reducing execution time and improving fidelity by leveraging SIMD-style transport in both intra- and inter-trap operations.
% (Resolved) % \textcolor{red}{(Jens: what does ``co-optimize'' mean?  This is the only place in the paper that I found this term and I am unsure what it means and what the sentence is trying to say.)!} 
% While our SIMD-style architecture and gate-zone-aware strategies are key to improving performance, t
This evaluation also provides practical insights for future hardware design, especially under varying architectural constraints and connectivity trade-offs.

\subsection{Experiment Setup}\label{subsect:experiment_setup}
\noindent \textbf{Baselines.} 
We compare our compiler against two representative baselines adapted to the QCCD architecture: SHAPER* and QCCD-SIM. 
SHAPER* is adapted from state-of-the-art SHAPER strategy~\cite{02_28_shaper}, originally developed for all-to-all connected architectures including superconducting and ion-trap systems. We reconstruct a baseline by combining its position graph abstraction with SABRE-style heuristics, tailored to our trapped-ion hardware constraints. 
% The SHAPER paper reports that its method consistently outperforms QCCD-SIM across benchmarks, supporting its use as our primary baseline.
QCCD-SIM~\cite{03_02_qccdsim} shares SHAPER’s idealized connectivity model but uses a simpler inter-trap scheduling strategy. It lacks a general scheduling framework and does not support intra-trap transport, so we evaluate its inter-trap transport component only, combined with our intra-trap solution. For fair comparison. 
% For fairness, we retain each baseline’s scheduling and mapping, but augment their transport layer 
we strengthen each baseline with a SIMD-style post pass that groups aligned transports in their scheduling into SIMD instructions, enabling parallelism absent in their native frameworks.
Further details on both baselines and their adaptations can be found in Section~\ref{sect:background}.

\begin{table}[h]
    \centering
      \caption{NISQ Benchmark Programs.}
    \resizebox{0.5\textwidth}{!}{
        \renewcommand*{\arraystretch}{0.95}
        \begin{footnotesize}
        \begin{tabular}{|m{0.7cm}|m{1.1cm}|m{1.8cm}|m{1.1cm}|m{1.8cm}|m{1.3cm}|}  \hline
         \#Qubits & DxD (Grid Shape) & Grid Layout & Number of 1D Trap & $N_{gz}$ (Number of Gate Zones / Trap)& L (1D Trap Capacity)\\ 
        \hline
         \multirow{6}{*}{20} & \multirow{6}{*}{2x2} & \multirow{6}{*}{\includegraphics[width=0.09\textwidth]{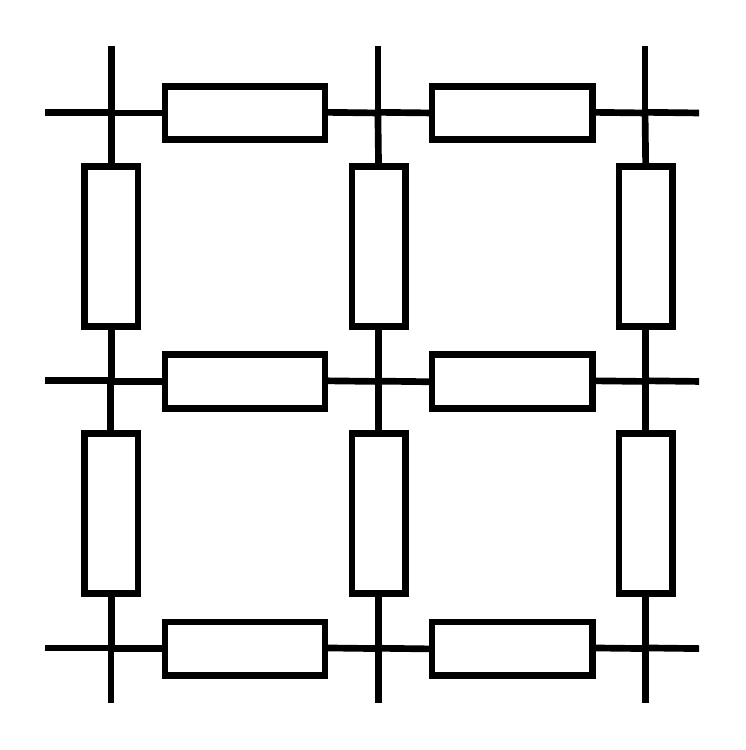}} & \multirow{6}{*}{12} & \multirow{6}{*}{2} & \multirow{3}{*}{8}\\
         &&&&&\\
         &&&&&\\
         \cline{6-6}&&&&& \multirow{3}{*}{14}\\
         &&&&&\\
         &&&&&\\
         \hline
         \multirow{6}{*}{40} & \multirow{6}{*}{3x3} & \multirow{6}{*}{\includegraphics[width=0.09\textwidth]{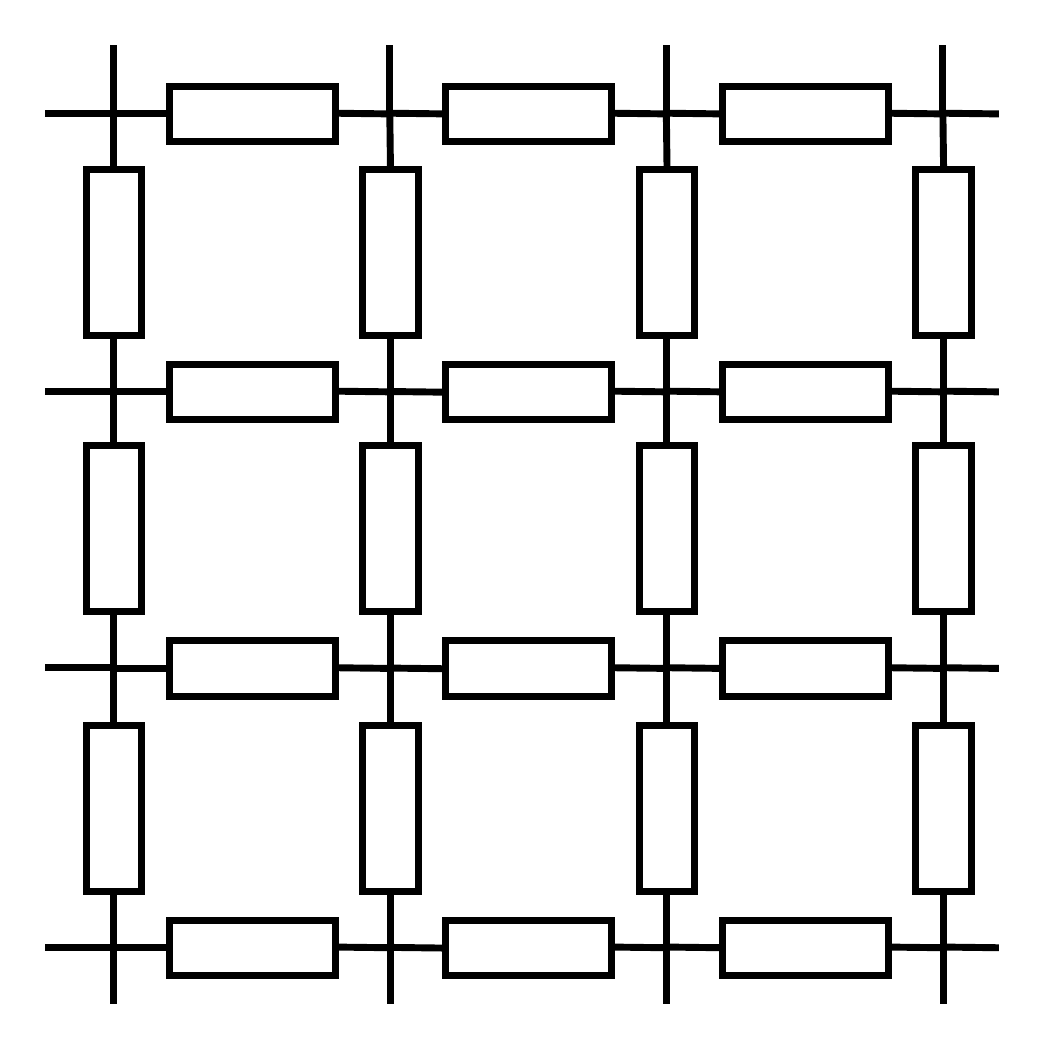}} & \multirow{6}{*}{24} & \multirow{6}{*}{2} & \multirow{3}{*}{8}\\
         &&&&&\\
         &&&&&\\
         \cline{6-6}&&&&& \multirow{3}{*}{14}\\
         &&&&&\\
         &&&&&\\
         \hline
         \multirow{6}{*}{60} & \multirow{6}{*}{4x4} & \multirow{6}{*}{\includegraphics[width=0.09\textwidth]{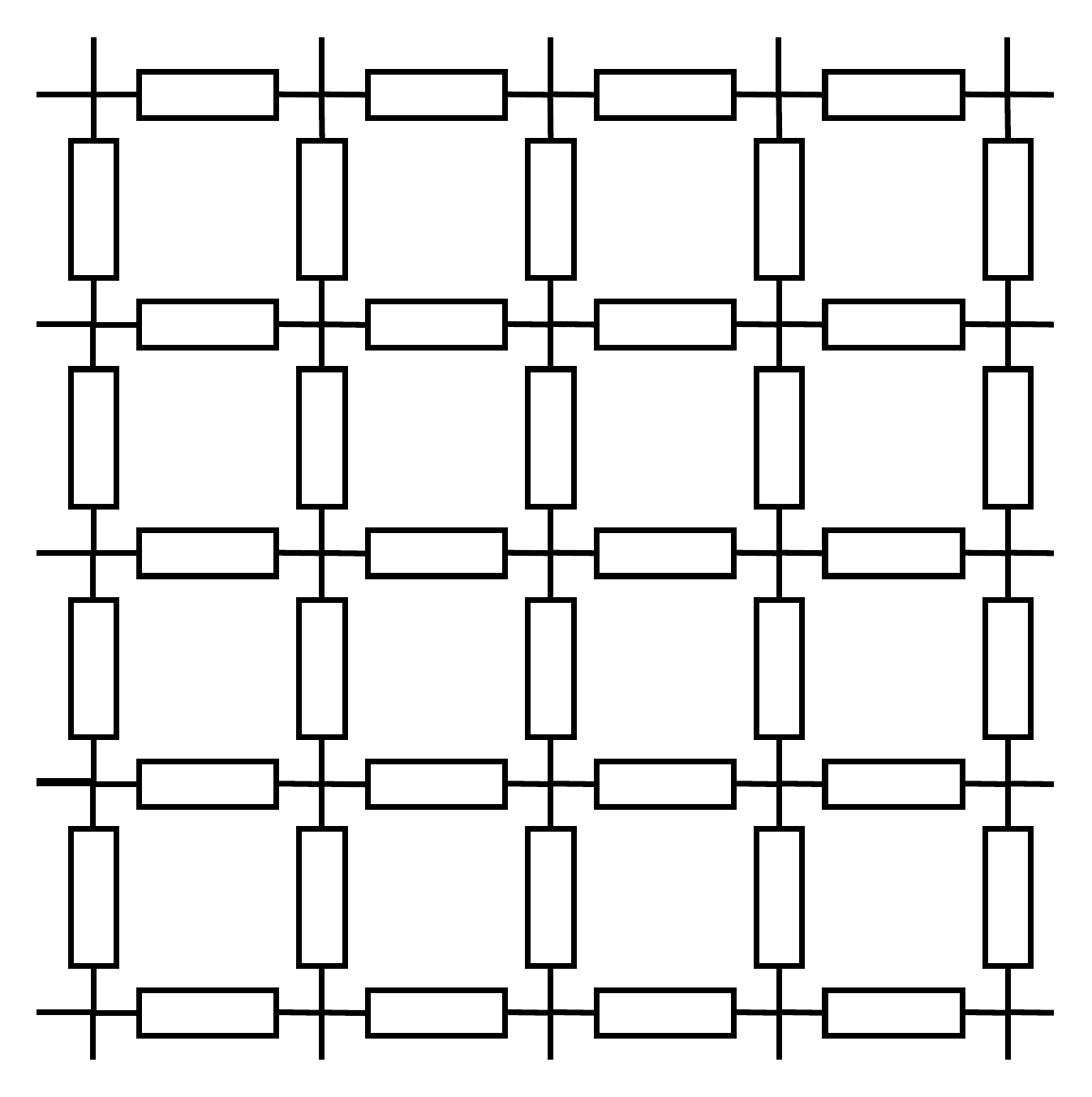}} & \multirow{6}{*}{40} & \multirow{6}{*}{2} & \multirow{3}{*}{8}\\
         &&&&&\\
         &&&&&\\
         \cline{6-6}&&&&& \multirow{3}{*}{14}\\
         &&&&&\\
         &&&&&\\
         \hline

         \multirow{6}{*}{100} & \multirow{6}{*}{5x5} & \multirow{6}{*}{\includegraphics[width=0.09\textwidth]{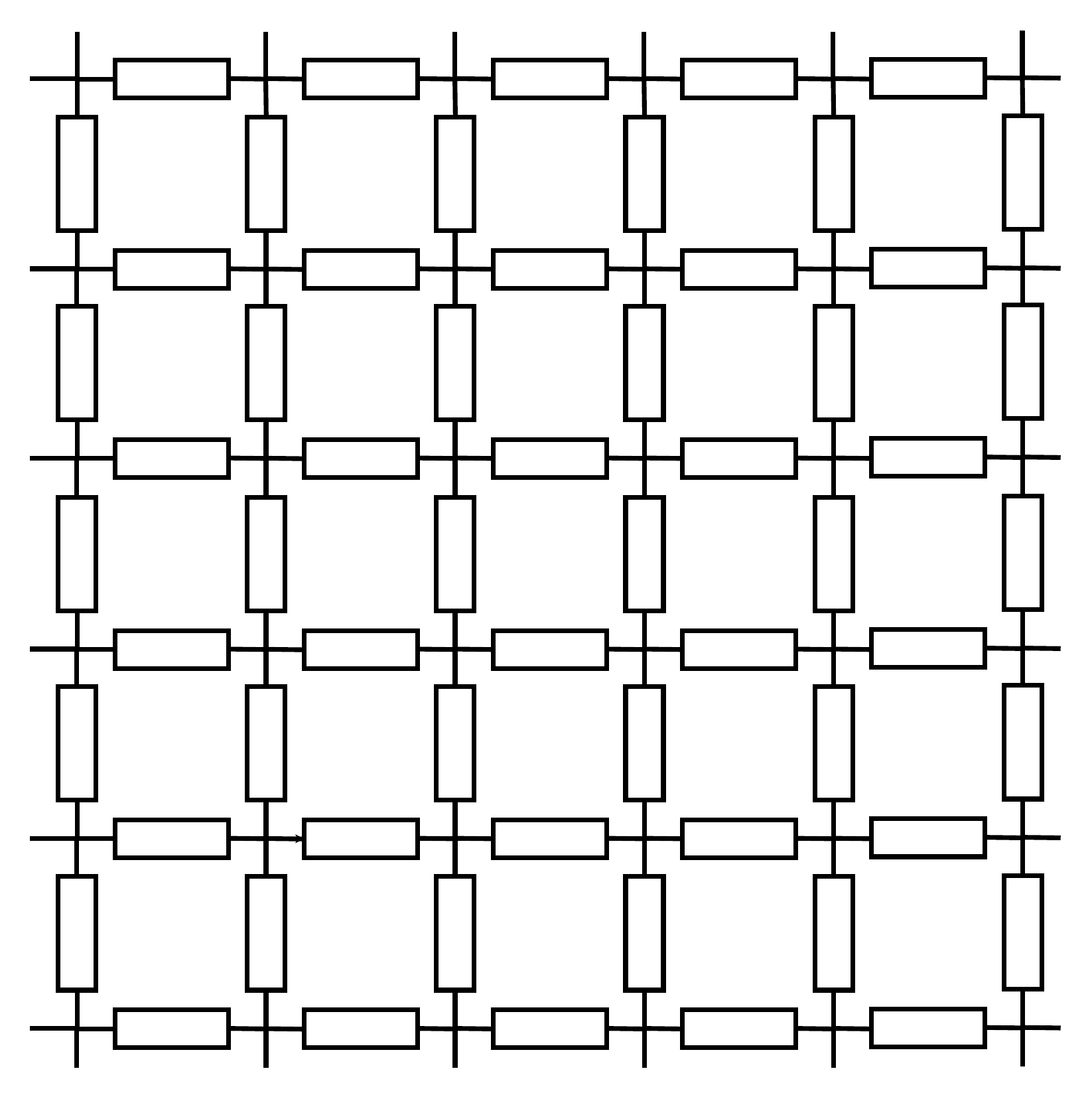}} & \multirow{6}{*}{60} & \multirow{6}{*}{2} & \multirow{3}{*}{8}\\
         &&&&&\\
         &&&&&\\
         \cline{6-6}&&&&& \multirow{3}{*}{14}\\
         &&&&&\\
         &&&&&\\
         \hline
        \end{tabular}
        \end{footnotesize}
    }
    \label{tab:benchmark}
\end{table}

\vspace{0.3em}
\noindent \textbf{Benchmarks.} We evaluate performance using a diverse suite of benchmarks under both NISQ and FTQC settings.

\noindent (1) \emph{NISQ Applications.} 
We select a variety of benchmark programs, including Quantum Approximate Optimization Algorithm (QAOA), Ripple-Carry Adder (RCA)~\cite{RCA}, Bernstein-Vazirani (BV), and Variational Quantum Eigensolver (VQE), across multiple circuit sizes, as summarized in Table~\ref{tab:benchmark}. QAOA is implemented on randomly generated 3-regular graphs, where ZZ gates are applied only between graph-connected qubits. BV circuits are generated using random secret strings with balanced distributions of 0s and 1s. For VQE, we employ a standard full-entanglement ansatz, known for its expressive power~\cite{qiskit_vqe, vqe_ansatz}.

\noindent (2) \emph{FTQC Applications.} 
To evaluate \frameworkname{} in the fault-tolerant regime, we use QFT-20 and QSim-20 \cite{qsim0, qsim1} circuits protected by surface codes, exploring code distances ranging from 3 to 11. This selection ensures realistic program success probabilities while maintaining hardware resource demands within the scope of expected advancements over the next 5–10 years (physical error rate: 0.1\%–0.01\%, total qubit count: 1,000–20,000) \cite{ft_ref0, ft_ref1}.

\vspace{0.3em}
\noindent \textbf{Metrics.} We evaluate our compiler’s performance using different metrics tailored to the NISQ and FTQC settings described above:

\noindent (1) \emph{NISQ Applications.} 
In comparisons with \textit{SHAPER*}, we consider two primary metrics: \emph{Execution Time} ($T_{\text{exe}}$), which includes the total duration of 1Q and 2Q gates as well as intra-/inter-trap shift and swap operations; and \emph{Fidelity (F)}, modeled as 
$F = F_{1Q} \cdot F_{2Q} \cdot F_{\text{transport}} \cdot F_{decoh}$ \cite{fidelity, 03_02_new_ion_hardware}, 
where 
$F_{1Q} = f_{1Q}^{N_{1Q}}$, 
$F_{2Q} = f_{2Q}^{N_{2Q}}$, 
$F_{\text{transport}}$ $ = f_{\text{intra\_shift}}^{N_{\text{intra\_shift}}} \cdot f_{\text{intra\_swap}}^{N_{\text{intra\_swap}}} \cdot f_{\text{inter\_shift}}^{N_{\text{inter\_shift}}} \cdot f_{\text{inter\_swap}}^{N_{\text{inter\_swap}}}$, and $F_{\text{decoh}} = exp(- n$ $T_{exe}$ / $T_{coh})$ with 
$n$ denoting the number of qubits in the program.
Here, $N$ terms denote the respective operation counts, while each fidelity term reflects the accumulated degradation from specific gate or transport operations. We set the coherence time $T_{coh}=600s$ based on reported values from state-of-the-art TI hardware~\cite{coherence}. Additionally, when comparing inter-trap scheduling performance against QCCDsim and SHAPER*, we report inter-trap transport schedules as a separate metric to account for differences in hardware instruction latency.

\noindent (2) \emph{FTQC Applications.} For fault-tolerant workloads, we focus on \emph{Execution Time} ($T_{\text{exe}}$) as the primary metric, since we can suppress the logical error rate by increasing the code distance \cite{ft_ref0}, as we show in Figure~\ref{fig:qec_result}(a).

\noindent \textbf{Hardware setting.}
For an $n$-qubit benchmark, the hardware layout is shown in Table~\ref{tab:benchmark}. Given that Quantinuum's 1D ion trap hardware can accommodate a maximum of around 20 ion qubits \cite{03_02_new_ion_hardware}, we selected trap sizes of 8 and 14 qubits. Each trap has 2 gate zones for two-qubit gates. We chose this layout with more vacant space because the maximum program size that SHAPER* can schedule is limited by the number of gate zones. Performance in more compact layouts will be discussed further in Figure~\ref{fig:sensitivity}.

\noindent \textbf{Evaluation Methodology.}  
This subsection details the evaluation methodology for the two experimental settings.

\noindent (1) \emph{NISQ Applications.}  
\frameworkname{} computes program execution time with the discrete event scheduler described in Section~\ref{sect:scheduler}, 
% which includes single-qubit and two-qubit gates, as well as intra-trap and inter-trap swaps and shifts, after applying parallelization and pipelining. The with the timing simulation 
based on instruction durations listed in Table~\ref{tab:hardware_data_table}.  To model fidelity, we compute the overall circuit fidelity using the total execution time and operation counts, i.e., $N_{1Q}$, $N_{2Q}$, $N_{\text{intra\_shift}}$, $N_{\text{intra\_swap}}$, $N_{\text{inter\_shift}}$, and $N_{\text{inter\_swap}}$, alongside their respective instruction fidelities, also listed in Table~\ref{tab:hardware_data_table}.

\noindent (2) \emph{FTQC Applications.} In the fault-tolerant setting, 
% \frameworkname\ schedules operations acting on 
operations are performed on entire surface-code logical qubits. Programs are decomposed to Clifford+T gate set \cite{bravyi2005universal},
% , with circuits decomposed into layers of CNOT, H, and T gates. 
with logical CNOT and H gates executed transversally
% , allowing parallelism across disjoint qubit pairs. The 
and logical T gates implemented via gate teleportation~\cite{bravyi2005universal, litinski2019magic, bravyi2012magic}.
% , which involves CNOT interactions and Clifford corrections. 
Unlike NISQ operations, the scheduling of T gates must also coordinate the transport of data qubits toward pre-generated magic states, introducing additional spatial and temporal complexity.
We adopt the surface-code layout mapping strategy from~\cite{02_25_ion_surface_code_compiler}, which maps each logical patch onto a grid of traps, with one physical qubit—either data or syndrome—assigned per trap to facilitate parallel control and scalable scheduling in QCCD architectures. 
% We compute execution time using the scaled layout and hardware timing model. 
To estimate logical error rates, we use \texttt{Stim} simulator~\cite{stim} under the TI noise profile shown in Table \ref{tab:hardware_data_table} and Pymatching for decoding and error correction~\cite{higgott2022pymatching}. In our experiments, we focus on the execution time of logical operations and omit syndrome extraction time, as it is roughly proportional to the logical gate latency~\cite{ft_ref0}.
\begin{figure}[!h]
        \centering
        \includegraphics[width=0.82\linewidth, trim=0.1in 0.1in 0.1in 0.1in, clip]{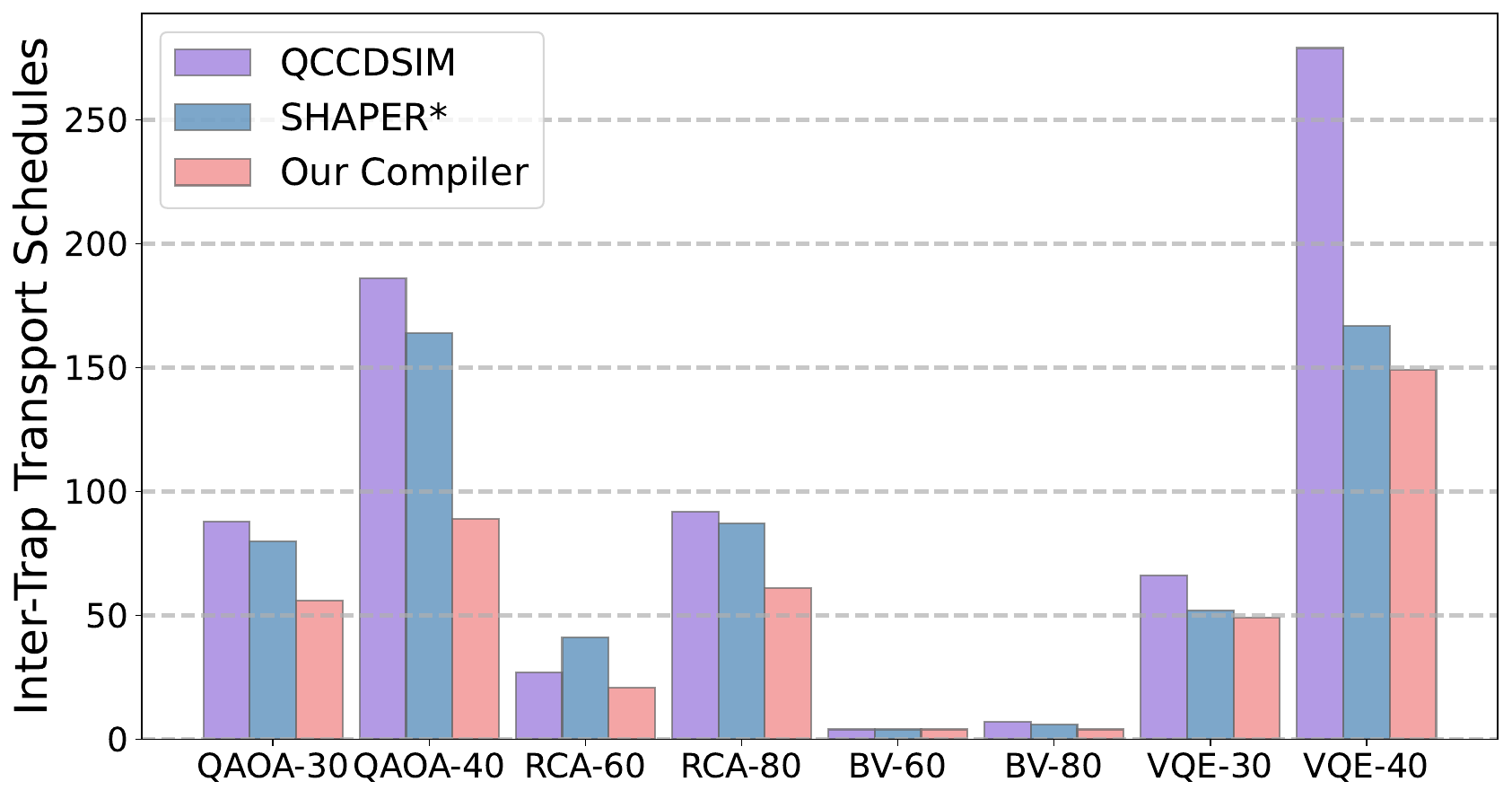}

        \caption{Comparison of Inter-Trap Transport Performance}
        \label{fig:inter_trap_transport_schedules}
\end{figure}

\subsection{Results for NISQ Applications}\label{subsection: main results}
We first compare \frameworkname~ with SHAPER* under two 1D trap capacities (\(N_{trap} = 8\) and \(N_{trap} = 14\)), focusing on execution time (\(T_{exe}\)) and fidelity. To better understand the source of improvements, we break down \(T_{exe}\) into intra- and inter-trap transport times, and fidelity into transport, decoherence and two-qubit gate fidelity. Secondly, We compare with QCCDsim focusing on inter-trap transport performance, as its routing strategy does not support intra-trap transport evaluation (Section~\ref{sect:background}).

\begin{table*}[]
    \centering
    \caption{Results of our compiler and SHAPER*. Each `benchmark-$n$' corresponds to an $n$-qubit circuit in the circuit model.
    % , compiled to a 2D size of  $\sqrt{n}\times\sqrt{n}$.
    }
  %  The name column are acronyms of test programs in Table~\ref{tab:benchmark}.}
    \resizebox{\textwidth}{!}{
        \renewcommand*{\arraystretch}{0.9}
        \begin{normalsize}
            \begin{tabular}{|p{1.5cm}||p{1.7cm}|p{1.7cm}|p{1.6cm}|p{1.8cm}||p{1.8cm}|p{1.8cm}|p{1.8cm}|}
        
        \hline
        L (1D Trap Capacity)  & Benchmark - \#Qubit & SHAPER* $T_{exe} (\mu s)$ & Our $T_{exe} (\mu s)$ & $T_{exe} (\mu s)$ \ \ \ \ \ \ \ \ \ \ \ \ \ \ Improv.  & SHAPER* \ \ \ \ \ \ \ \ \ \ \ \ \ \ Fidelity & Our Fidelity & Fidelity Improv.\\
        \hline
        \hline\multicolumn{1}{|c||}{\multirow{12}{*}{\parbox{2cm}{ 8 }}}
        & QAOA-20 & 173,561 & 93,845 & 1.85 & 0.49 & 0.61 & 1.25 \\
        \cline{2-8}
        & RCA-20 & 214,369 & 70,624 & 3.04 & 0.43 & 0.57 & 1.32 \\
        \cline{2-8}
        & BV-20 & 9,870 & 8,373 & 1.18 & 0.97 & 0.97 & 1.0 \\
        \cline{2-8}
        & VQE-20 & 218,974 & 130,538 & 1.68 & 0.42 & 0.56 & 1.33 \\
        \cline{2-8}
        & QAOA-40 & 733,637 & 327,601 & 2.24 & 0.03 & 0.13 & 4.21 \\
        \cline{2-8}
        & RCA-40 & 487,001 & 180,950 & 2.69 & 0.16 & 0.28 & 1.78 \\
        \cline{2-8}
        & BV-40 & 21,554 & 16,175 & 1.33 & 0.92 & 0.94 & 1.02 \\
        \cline{2-8}
        & VQE-40 & 883,977 & 532,440 & 1.66 & 0.02 & 0.09 & 4.66 \\
        \cline{2-8}
        & QAOA-60 & 1,684,718 & 744,943 & 2.26 & $2.93 \times 10^{-04}$ & $8.91 \times 10^{-03}$ & 30.36 \\
        \cline{2-8}
        & RCA-60 & 801,437 & 278,970 & 2.87 & 0.05 & 0.14 & 2.82 \\
        \cline{2-8}
        & BV-60 & 37,866 & 23,625 & 1.6 & 0.88 & 0.92 & 1.05 \\
        \cline{2-8}
        & VQE-60 & 1,921,069 & 1,255,877 & 1.53 & $8.54 \times 10^{-05}$ & $3.33 \times 10^{-03}$ & 39.04 \\
        \cline{2-8}
        & QAOA-100 & 4,687,884 & 2,256,313 & 2.08 & $3.56 \times 10^{-11}$ & $5.85 \times 10^{-07}$ & $1.64 \times 10^{4}$ \\
        \cline{2-8}
        & RCA-100 & 1,346,070 & 561,586 & 2.4 & $5.25 \times 10^{-03}$ & 0.03 & 5.65 \\
        \cline{2-8}
        & BV-100 & 66,858 & 38,679 & 1.73 & 0.74 & 0.86 & 1.16 \\
        \cline{2-8}
        & VQE-100 & 5,200,261 & 3,570,854 & 1.46 & $8.33 \times 10^{-13}$ & $8.88 \times 10^{-08}$ & $1.07 \times 10^{5}$ \\
        \hline
        \multicolumn{1}{|c||}{\multirow{12}{*}{\parbox{2cm}{ 14 }}}
        & QAOA-20 & 265,068 & 128,886 & 2.06 & 0.38 & 0.58 & 1.53 \\
        \cline{2-8}
        & RCA-20 & 303,805 & 85,702 & 3.54 & 0.35 & 0.56 & 1.63 \\
        \cline{2-8}
        & BV-20 & 13,233 & 11,745 & 1.13 & 0.96 & 0.97 & 1.01 \\
        \cline{2-8}
        & VQE-20 & 300,942 & 134,186 & 2.24 & 0.3 & 0.55 & 1.86 \\
        \cline{2-8}
        & QAOA-40 & 1,053,797 & 370,858 & 2.84 & $8.02 \times 10^{-03}$ & 0.12 & 15.23 \\
        \cline{2-8}
        & RCA-40 & 689,769 & 204,547 & 3.37 & 0.1 & 0.27 & 2.76 \\
        \cline{2-8}
        & BV-40 & 31,426 & 18,479 & 1.7 & 0.9 & 0.94 & 1.05 \\
        \cline{2-8}
        & VQE-40 & 1,223,741 & 784,346 & 1.56 & $3.07 \times 10^{-03}$ & 0.07 & 21.41 \\
        \cline{2-8}
        & QAOA-60 & 2,403,532 & 822,641 & 2.92 & $1.25 \times 10^{-05}$ & $8.01 \times 10^{-03}$ & 639.78 \\
        \cline{2-8}
        & RCA-60 & 1,131,689 & 296,592 & 3.82 & 0.02 & 0.14 & 6.74 \\
        \cline{2-8}
        & BV-60 & 50,583 & 28,741 & 1.76 & 0.81 & 0.91 & 1.13 \\
        \cline{2-8}
        & VQE-60 & 2,669,385 & 1,551,403 & 1.72 & $1.05 \times 10^{-06}$ & $2.48 \times 10^{-03}$ & 2357.99 \\
        \cline{2-8}
        & QAOA-100 & 6,450,674 & 2,568,088 & 2.51 & $1.19 \times 10^{-15}$ & $3.51 \times 10^{-07}$ & $2.94 \times 10^{8}$ \\
        \cline{2-8}
        & RCA-100 & 1,900,434 & 572,560 & 3.32 & $1.06 \times 10^{-03}$ & 0.03 & 29.59 \\
        \cline{2-8}
        & BV-100 & 106,961 & 60,293 & 1.77 & 0.66 & 0.83 & 1.27 \\
        \cline{2-8}
        & VQE-100 & 7,226,433 & 4,329,181 & 1.67 & $1.20 \times 10^{-18}$ & $9.58 \times 10^{-09}$ & $8.00 \times 10^{9}$ \\
        \hline
    \end{tabular}
    \end{normalsize}
    }
    \label{tab:main_table}
\end{table*}

% \vspace{2pt}

\begin{figure*}[]
        \centering
        \includegraphics[width=0.24\linewidth]{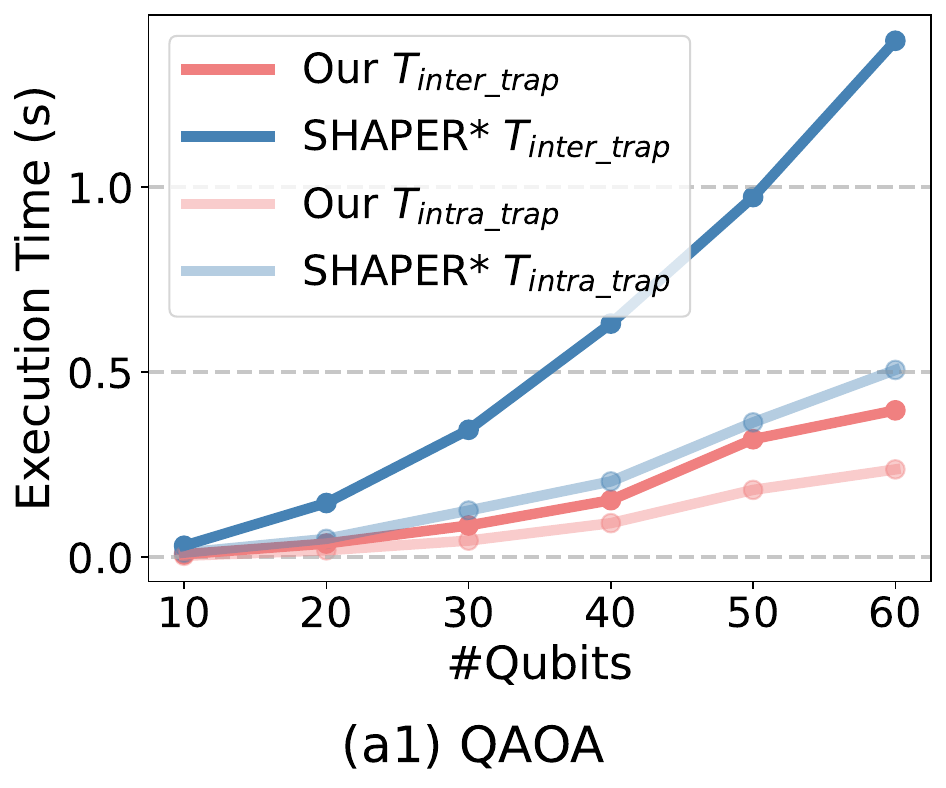} 
        \includegraphics[width=0.24\linewidth]{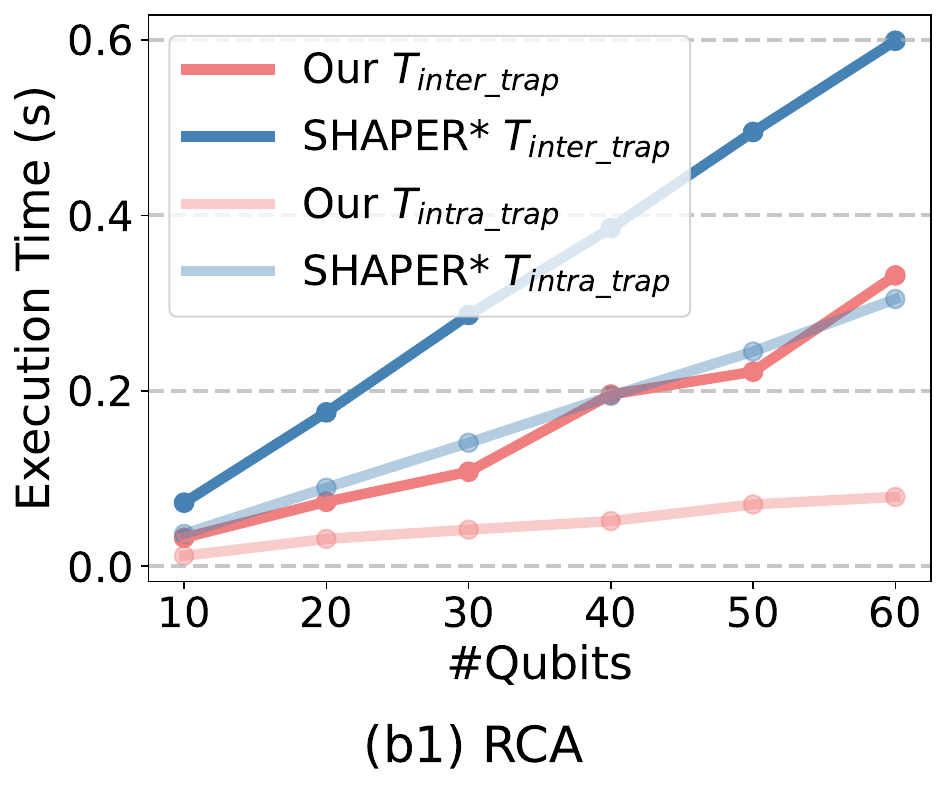}
        \includegraphics[width=0.24\linewidth]{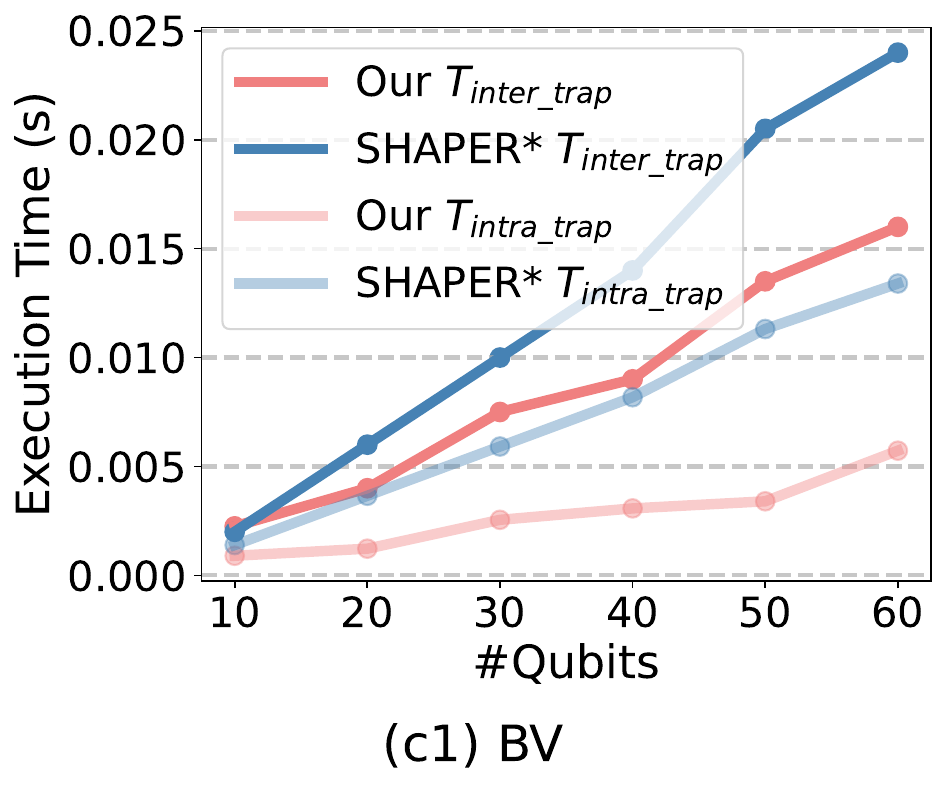}
        \includegraphics[width=0.24\linewidth]{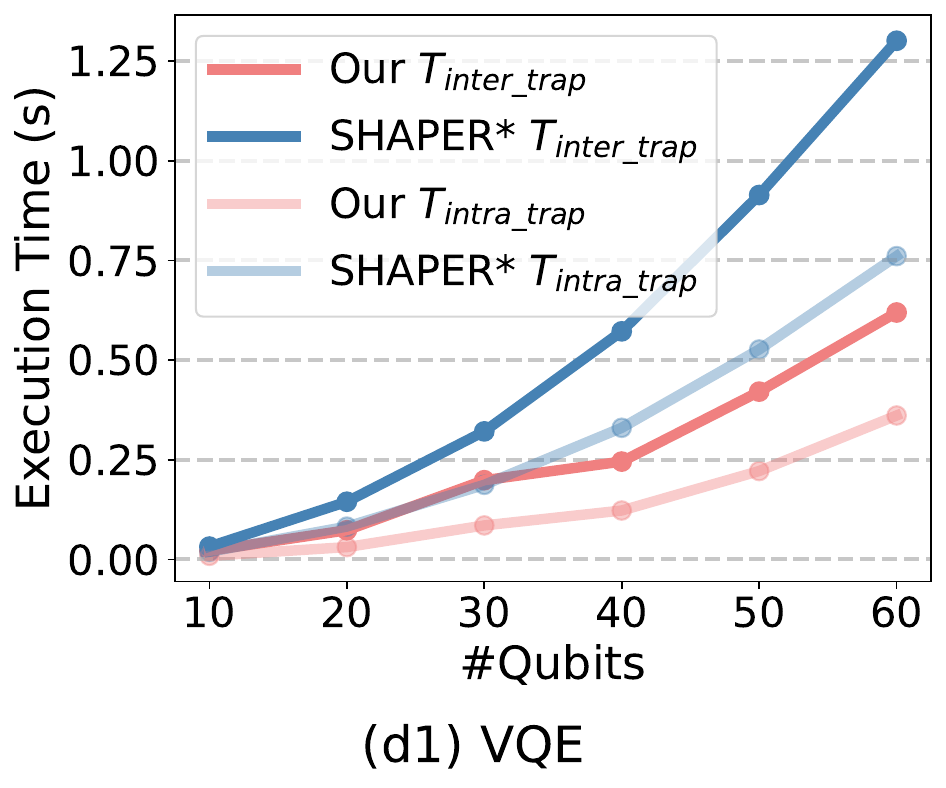}\\
        \includegraphics[width=0.24\linewidth]{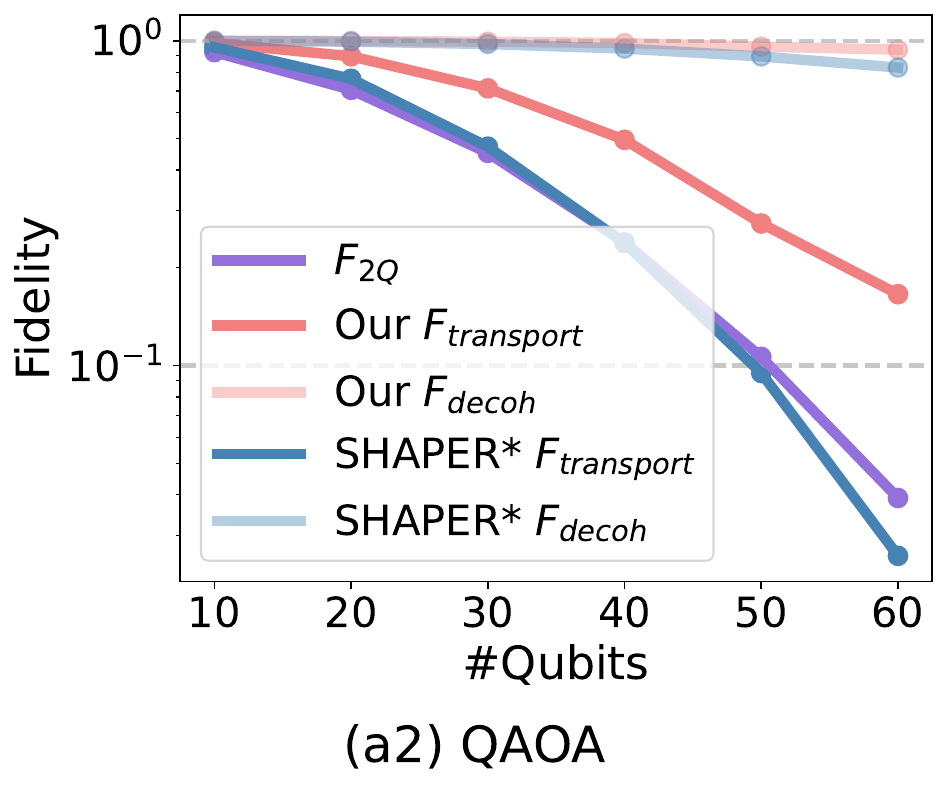} 
        \includegraphics[width=0.24\linewidth]{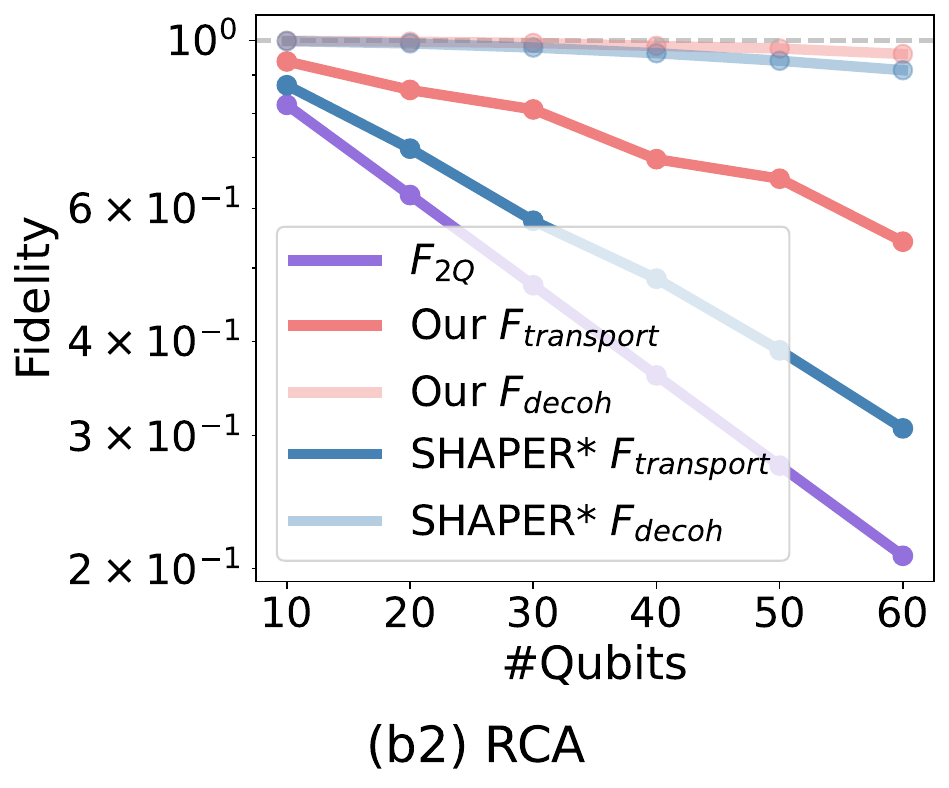}
        \includegraphics[width=0.24\linewidth]{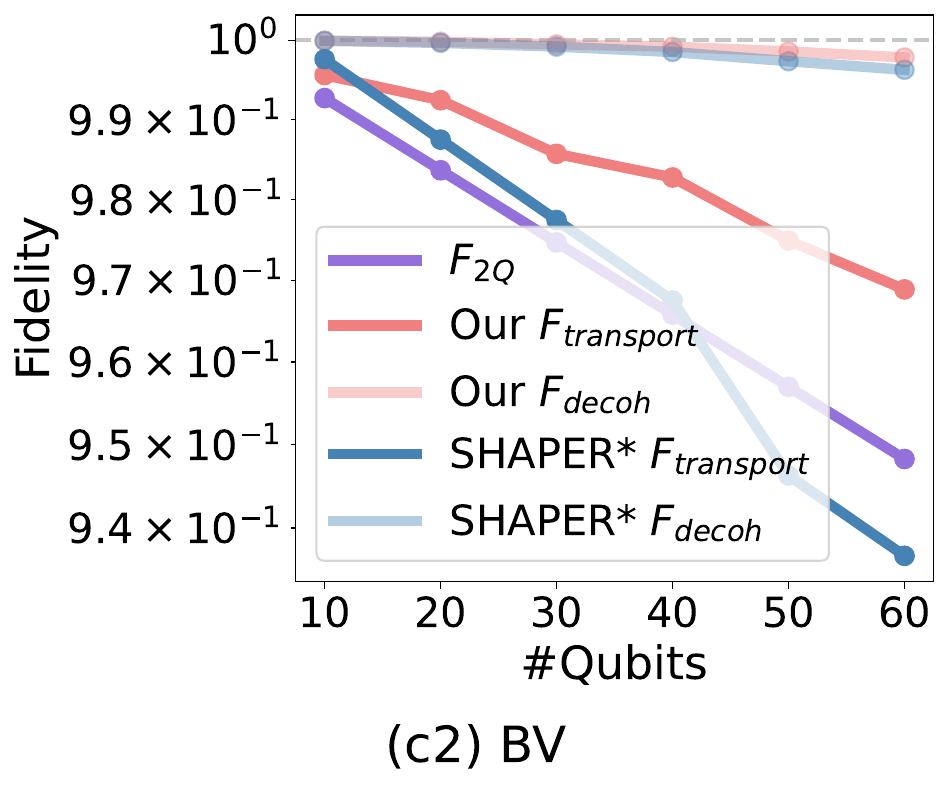}
        \includegraphics[width=0.24\linewidth]{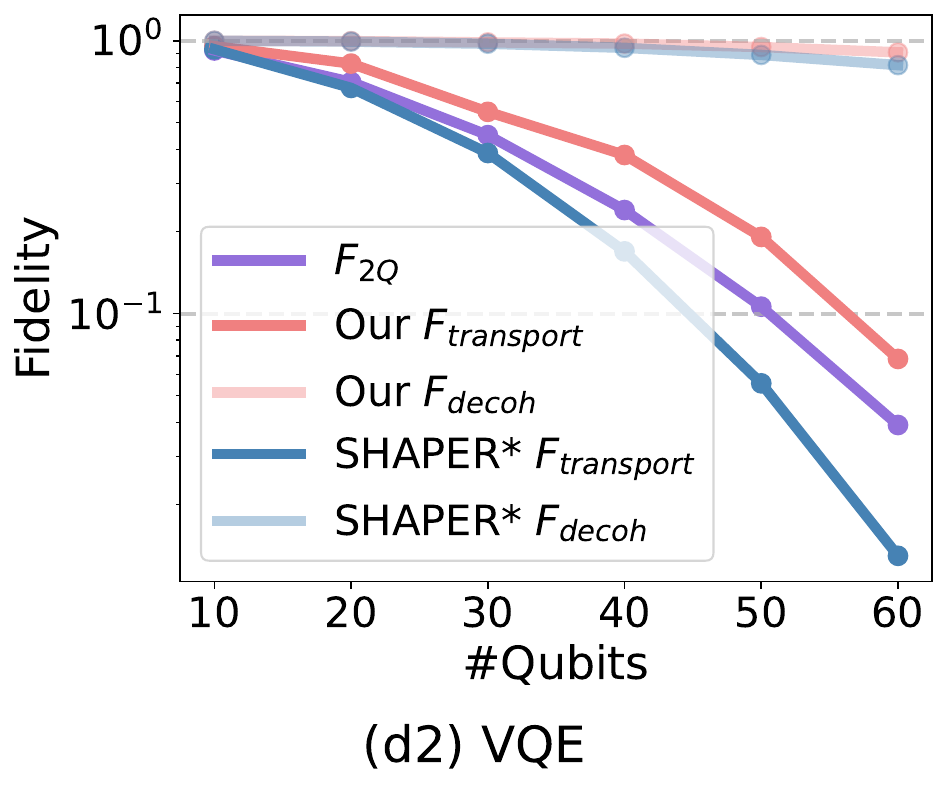}\\
        % (a)\hspace{115pt}(b)\hspace{112pt}
        % (c)\hspace{115pt}(d) 
        % \caption{}
        \caption{(a1)–(d1) show execution time breakdowns for QAOA, RCA, BV, and VQE, highlighting intra- vs. inter-trap transport. (a2)–(d2) show fidelity breakdowns into two-qubit gate, transport (shift/swap) errors and decoherence errors.}

       \label{fig:ablation}
\end{figure*}

\vspace{0.3em}
\noindent \textbf{Overall Performance.}
Table~\ref{tab:main_table} shows that \frameworkname~ significantly outperforms SHAPER* in both execution time and fidelity. For $L = 8$, we observe up to a \textbf{3.04$\times$} reduction in execution time and multi-order-of-magnitude improvements in fidelity across benchmarks of varying sizes. At $L = 14$, these improvements grow further—reaching up to \textbf{3.82$\times$} in execution time and even greater gains in fidelity. These results underscore the scalability and efficiency of our compiler, enabled by its SIMD-aware design that promotes coordinated routing and execution via aggregated transport.

\vspace{0.3em}
\noindent \textbf{Ablation Analysis.}
We further compare the breakdown of time and fidelity components with SHAPER* on a 5$\times$5 layout with 1D trap capacity of 3 and one computation zone per trap. We adopt this constrained configuration to amplify the scheduling challenges and better highlight the benefits of our SIMD-aware transport and coordination strategies. The ablation study helps isolate the benefits introduced by our SIMD-aware instruction set and scheduling strategy.

\noindent (1) \emph{Time Ablation.}
Figure~\ref{fig:ablation}(a1)–(d1) break down intra- and inter-trap execution times, with our results shown in red and SHAPER* in blue. In each plot, the solid line represents inter-trap transport time, while the lighter line denotes intra-trap transport. Although inter-trap transport generally dominates total latency due to its higher per-operation cost, intra-trap movement remains non-negligible, especially in interaction-heavy circuits. Our compiler consistently reduces both components, with particularly notable improvements in QAOA and VQE benchmarks that involve more remote interactions, as shown in Figure~\ref{fig:ablation}(a1)(d1). These gains stem from our SIMD-aware design: S3 instructions accelerate intra-trap shifts via parallel execution, while JT-SIMD batching reduces synchronization overhead for inter-trap transport. This is further enhanced by our gate zone-aware heuristic, which aligns qubit transport and aggregation patterns with zone availability to minimize routing cost. The advantages become more pronounced with increasing qubit count, highlighting the scalability of our approach.

\noindent (2) \emph{Fidelity Ablation.}
Figure~\ref{fig:ablation}(a2)–(d2) shows a fidelity breakdown comparing SHAPER* (blue/light blue) and \frameworkname~ (red/light red), with the common two-qubit gate fidelity plotted as a purple reference line. 
% Generally, transport operations outperform two-qubit gates in fidelity, and decoherence effects remain minimal. 
Intuitively, the overall transport fidelity should outperform that of two-qubit gates, as each transport operation has significantly higher fidelity than two-qubit gates.
% However, inefficient scheduling in SHAPER* introduces excessive transport and long execution time, pushing transport fidelity below the gate baseline and amplifying decoherence loss.
However, under inefficient scheduling strategies such as SHAPER*, excessive ion transport can degrade overall transport fidelity, pushing it below that of two-qubit gates (solid blue in Figure~\ref{fig:ablation}(a1)–(d1)). Additionally, prolonged execution time exacerbates decoherence errors (light red in Figure~\ref{fig:ablation}(a1)–(d1)).
In contrast, our SIMD-aware, hardware-compatible compiler minimizes redundant movements and shortens execution duration, preserving transport fidelity above the two-qubit gate baseline and keeping decoherence errors low—particularly in complex benchmarks like QAOA and VQE, as shown in Figure~\ref{fig:ablation}(a2)–(d2).

\vspace{0.3em}
\noindent \textbf{Inter-trap Scheduling Comparison with QCCDsim.}  
As shown in Figure \ref{fig:inter_trap_transport_schedules}, SHAPER* (blue) generally outperforms QCCDsim (purple) due to its more refined heuristic score analysis for routing steps. Notably, \frameworkname~ (red) achieves better performance than both, as it enhances routing with our SIMD-aware JT-SIMD transport model, which enables coordinated inter-trap transport and reduces synchronization overhead. On average, our approach delivers a $1.55\times$ improvement over QCCDsim and a $1.42\times$ improvement over SHAPER*. As the program size and routing complexity increase, the advantages of our method become more pronounced, particularly in benchmarks QAOA and VQE.

\subsection{Results for FTQC Applications}\label{subsection: extend_to_ftqc}

\begin{figure}[h]
        \centering
        \includegraphics[width=0.7\linewidth, trim=0.1in 0.1in 0in 0.1in, clip]{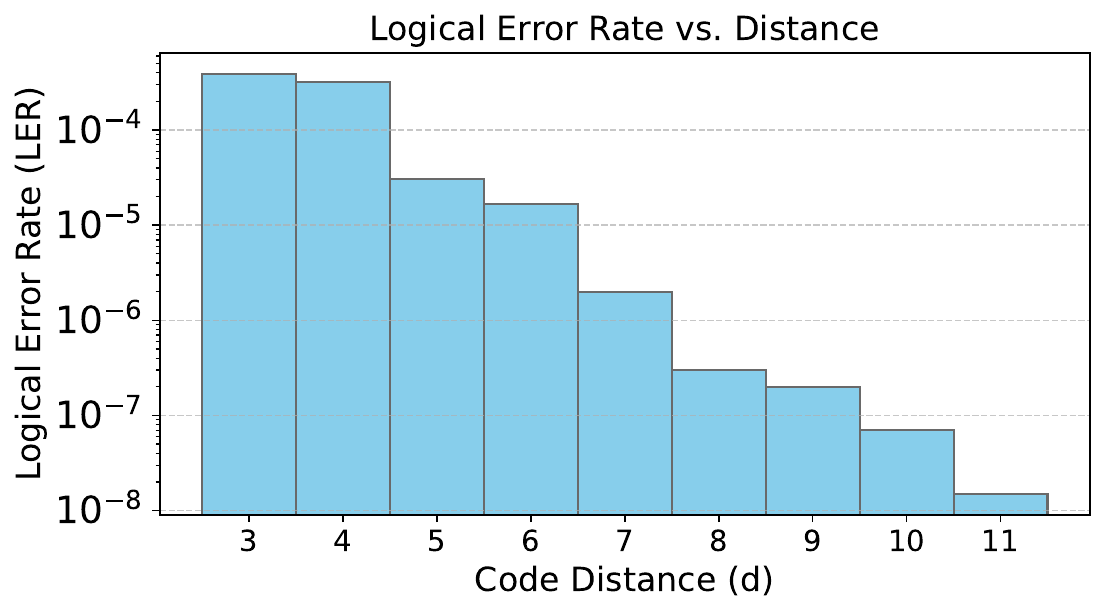}\hspace{120pt}(a)\\
        \includegraphics[width=0.48\linewidth]{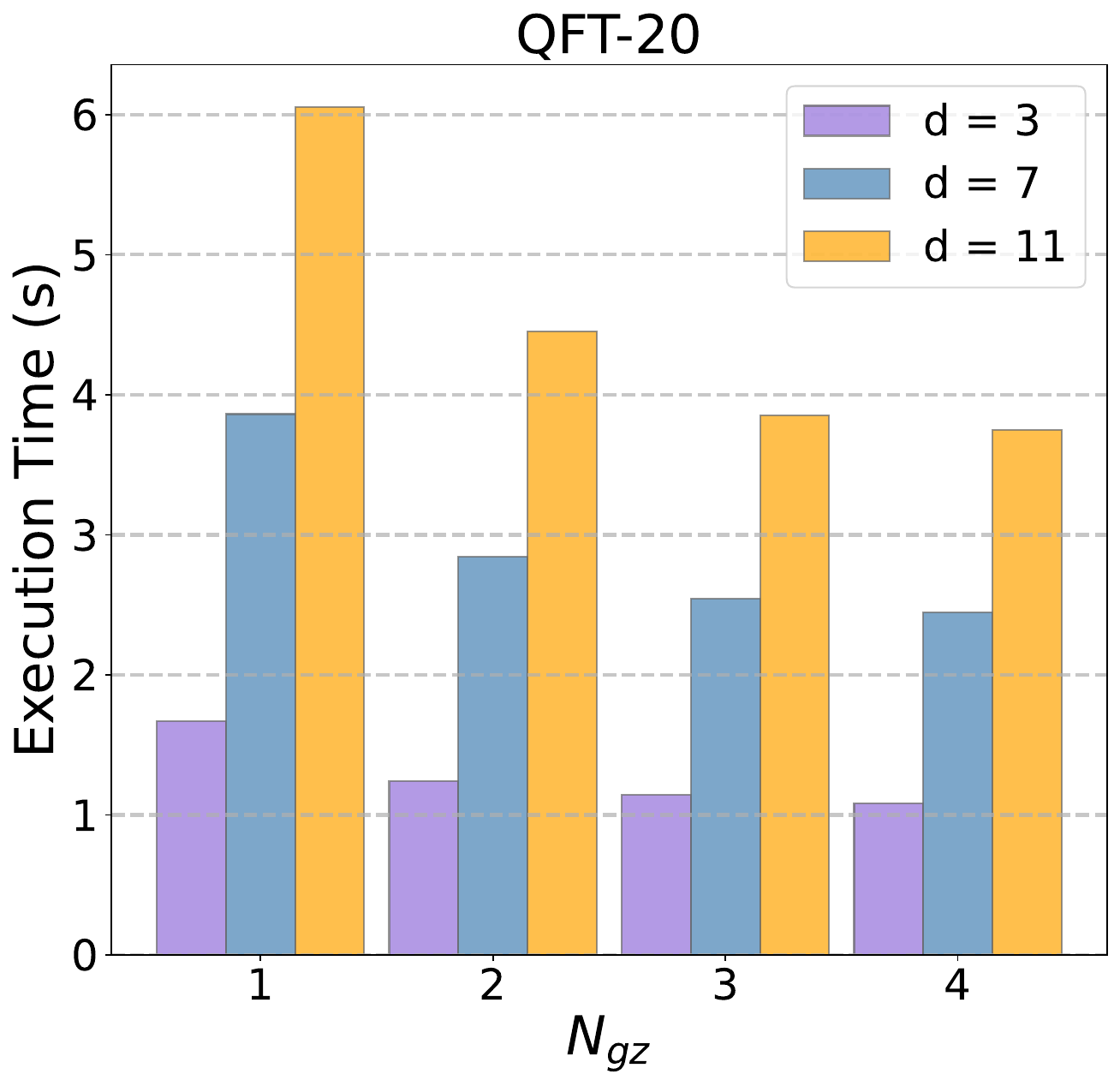}
        \includegraphics[width=0.49\linewidth]{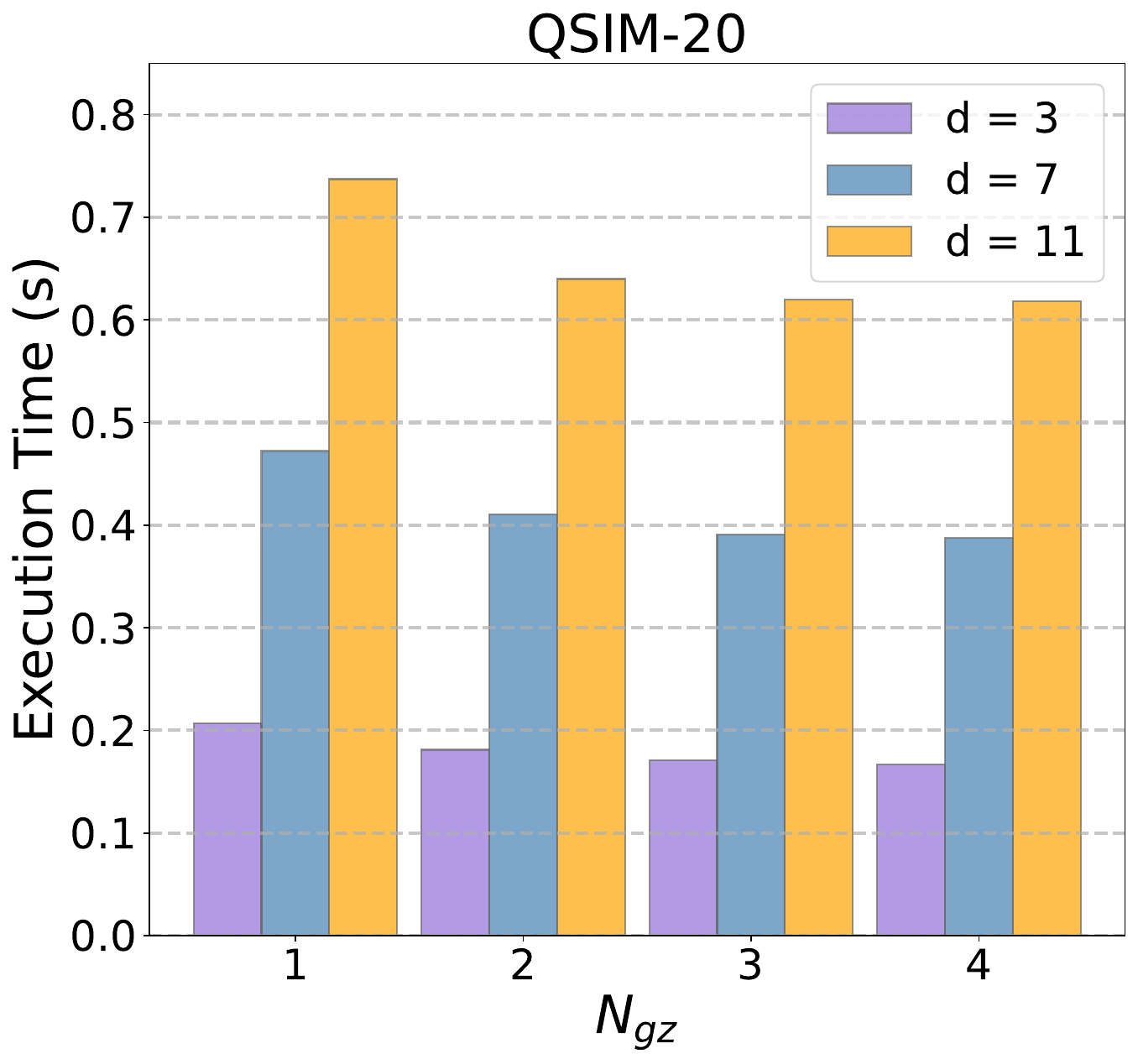} \\\hspace{20pt}(b)\hspace{110pt}(c)

        \caption{Evaluating Fault-Tolerant Execution: Logical Error Rates and Execution Time of QFT-20 and QSIM-20 under Varying Gate Zone Counts}

        \label{fig:qec_result}
\end{figure}

In this subsection, we present the results of extending our compiler to FTQC, with a focus on surface codes. First, we evaluate the LER for surface codes with varying distances under the physical error profile in Table \ref{tab:hardware_data_table}. As shown in Figure \ref{fig:qec_result}(a), a surface code with a distance of 11 achieves a logical error rate of \(10^{-8}\), which meets the LER requirement of our benchmarks. Next, we analyze the impact of gate zone density on execution time for surface codes with distances of 3, 7, and 11, using the QFT-20 and QSIM-20 benchmarks. As shown in Figures \ref{fig:qec_result}(b) and (c), a moderate number of zones can effectively reduce execution time, making a large number of gate zones unnecessary for realizing FTQC.

\subsection{Implications for TI Hardware Design}
To better understand how our SIMD-aware compiler adapts to different hardware settings, we conduct a sensitivity analysis under varying computation zone densities and hardware topologies, while keeping the total number of qubits fixed. Beyond demonstrating the flexibility of our approach, these experiments provide actionable insights into how hardware parameters—such as gate zone layout and trap granularity—affect system-level performance. These results reflect not only the benefits of SIMD-driven scheduling but also the effectiveness of our gate-zone-aware compilation strategy, offering practical guidance for co-designing future trapped-ion architectures.

\begin{figure}[!h]
        \centering
        \includegraphics[width=0.49\linewidth]{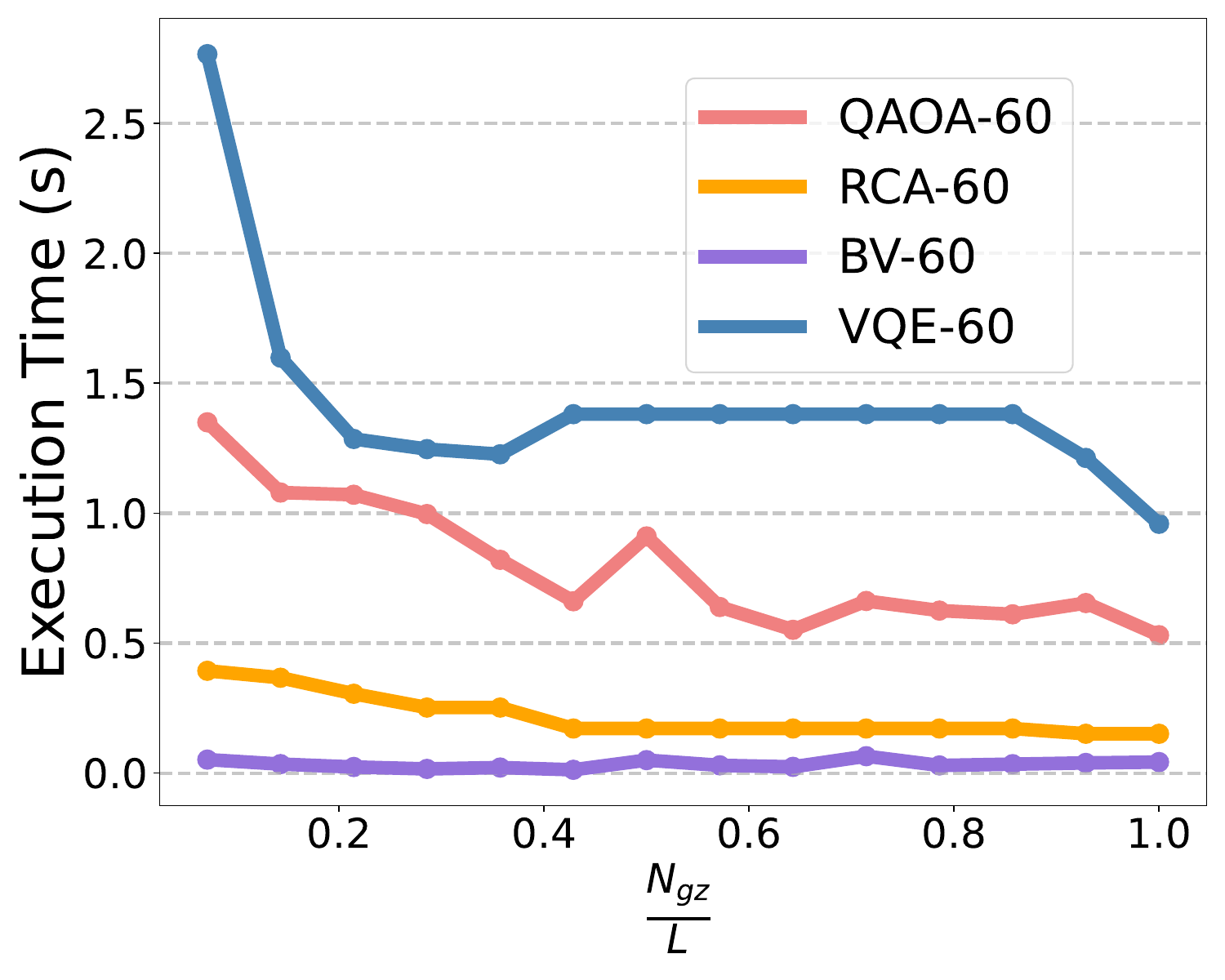}
        \includegraphics[width=0.49\linewidth]{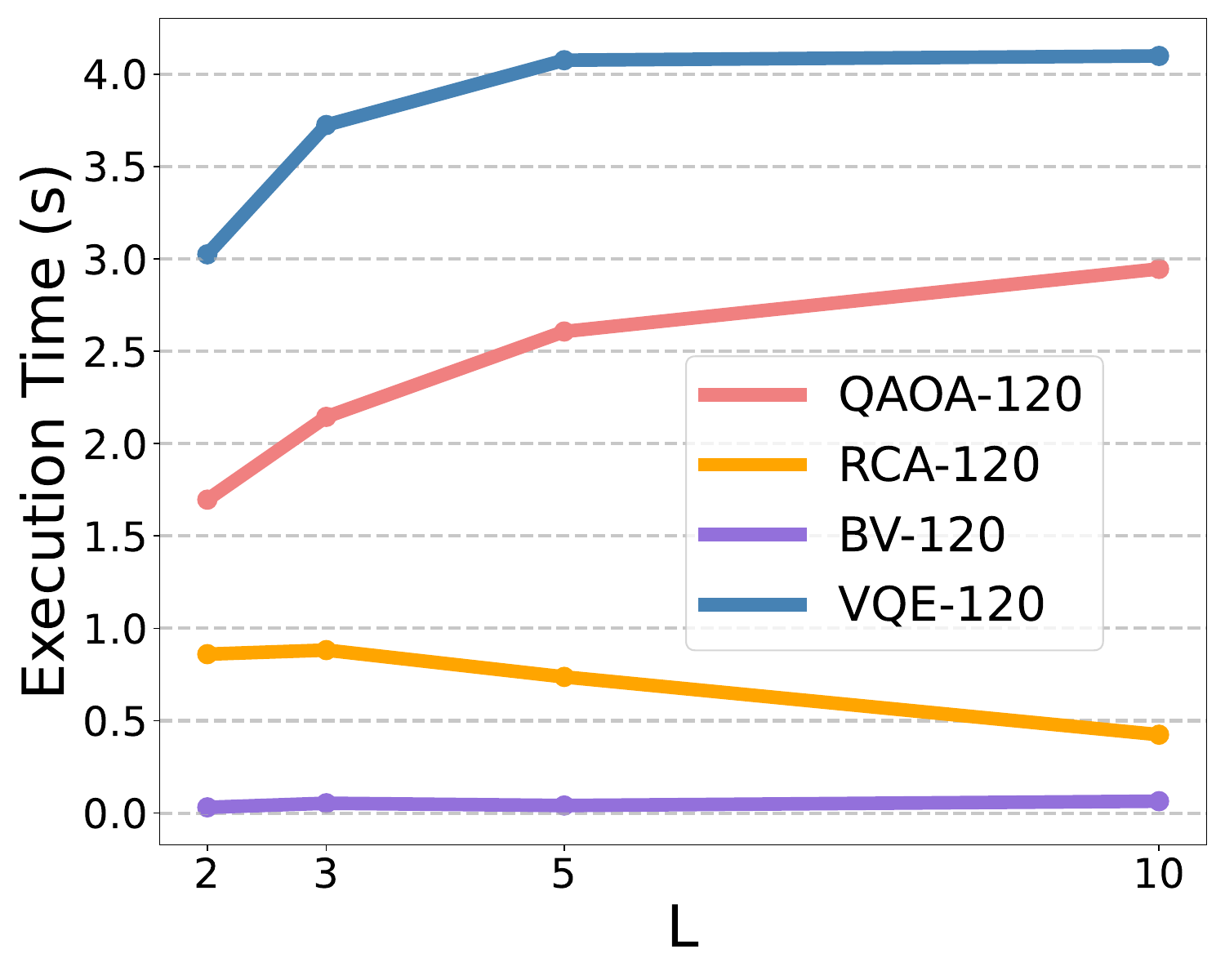} \\\hspace{20pt}(a)\hspace{110pt}(b)

        \caption{Sensitivity analysis under varying hardware parameters. $N_{gz}$ denotes the number of gate zones per 1D trap; $L$ denotes the 1D trap capacity.}
        \label{fig:sensitivity}
\end{figure}

\vspace{0.3em}
\noindent \textbf{Gate Zone Density.}
We evaluate the impact of gate zone density on execution time using a 3$\times$3 grid, where each 1D trap has a capacity of 14. As shown in Figure~\ref{fig:sensitivity}(a), execution time decreases as gate zone density increases, with improvements becoming marginal beyond 20--30\%, especially in the VQE benchmark. This suggests that over-provisioning computation zones brings limited returns. The optimal density is workload-dependent, and thus this result highlights an important hardware insight: gate zone allocation should be tuned based on program gate density rather than maximized blindly.

\vspace{0.3em}
\noindent \textbf{Impact of Hardware Topology Connectivity.}
We further explore how grid connectivity influences performance by varying 1D trap capacities under a fixed hardware budget of 120 qubits. Benchmarks are run on 5$\times$5, 4$\times$4, 3$\times$3, and 2$\times$2 grids corresponding to trap capacities of 2, 3, 5, and 10. As shown in Figure~\ref{fig:sensitivity}(b), benchmarks requiring highly non-local connectivity (e.g., QAOA and VQE) benefit from denser grid structures, which provide more flexible connectivity for routing. In contrast, RCA—with less frequent non-local interactions—performs better in sparser grids. These findings suggest that hardware designers should balance trap capacity and spatial density according to the expected workload class, rather than pursuing a one-size-fits-all design.

\section{Conclusion}
We present \frameworkname, a SIMD-aware compiler framework for modular trapped-ion quantum machines. By aligning compiler logic with the native transport behavior of QCCD architectures, FluxTrap enables efficient coordination of ion movement and gate execution. Its design combines a hardware-native SIMD abstraction with a scalable scheduling strategy, significantly improving execution time and fidelity. Evaluations across both NISQ and FTQC benchmarks confirm its effectiveness under realistic hardware constraints.
As trapped-ion systems scale toward tile-based architectures and memory-like resource models, we envision that FluxTrap’s SIMD abstraction can serve as a hardware-software contract, paving the way for portable, modular, and architecture-aligned quantum compilation.

\section{Acknowledgement}
This work is supported in part by NSF 2048144, NSF 2422169, and NSF 2427109.
Additional support was provided by NSF 2435382, Accelerating Fault-Tolerant Quantum Logic (FTL), and NSF 2016245, Quantum Leap Challenge Institute for Present and Future Quantum Computing (CIQC).
This work was also supported by the U.S. Department of Energy, Office of Science, National Quantum Information Science Research Centers, Quantum Science Center (QSC). This research used resources of the Oak Ridge Leadership Computing Facility, a DOE Office of Science User Facility supported under Contract DE-AC05-00OR22725.


% \bibliographystyle{ACM-Reference-Format}
\bibliographystyle{unsrturl}
\bibliography{references}

\end{document}